\documentclass[12pt]{article}
\usepackage[numbers,sort&compress]{natbib}
\usepackage{hyperref}
\hypersetup{breaklinks=true}
\usepackage{amsmath}
\usepackage{xcolor}
\RequirePackage{doi} 
\usepackage{authblk} 
\usepackage{multirow}
\usepackage{slashed}
\usepackage{float}
\usepackage{rotating}
\usepackage{caption}
\usepackage{breakcites}
\usepackage{microtype}
\usepackage[title]{appendix}
\graphicspath{ {./images/} }

\newcommand{\Amp}{\mathcal{M}}
\newcommand{\Ham}{\mathcal{H}}
\newcommand{\Ampl}{\mathcal{A}}
\newcommand{\mx}{\mathcal{F}}
\newcommand{\cng}[1]{\textcolor{black}{#1}}

\newcommand\ptwiddle[1]{\mathord{\mathop{#1}\limits^{\scriptscriptstyle(\sim)}}}
\newcommand{\mycomment}[2]{#2}

\title{New physics search via CP observables in $B^0_s \rightarrow \phi\phi$ decays with left- and right-handed Chromomagnetic operators}

\date{}

\author{Tejhas Kapoor\thanks{\href{mailto:tejhas.kapoor@ijclab.in2p3.fr}{tejhas.kapoor@ijclab.in2p3.fr}} }
\author{Emi Kou}
\affil{Universit\'e Paris-Saclay, CNRS/IN2P3, IJCLab, 91405 Orsay, France}

\begin{document}

\maketitle

\abstract In this paper, we investigate the time-dependent angular analysis of $B_s^0 \rightarrow \phi \phi$ decay to search for new physics signals via CP-violating observables.  We work with a new physics Hamiltonian containing both left- and right-handed Chromomagnetic dipole operators. The hierarchy of the helicity amplitudes in this model gives us a new scheme of experimental search, which is different from the ones LHCb has used in its analysis. To illustrate this new scheme, we perform a sensitivity study using two pseudo datasets generated using LHCb's measured values. We find the sensitivity of CP-violating observables to be of the order of $5-7\%$ with the current LHCb statistics. Moreover, we show that Belle(II)'s $B^0_d \rightarrow \phi K_s$ and LHCb's $B_s^0 \rightarrow \phi \phi$ measurements could be coupled within our model  to obtain the chirality of the new physics.


\newpage
\section{Introduction}
{Currently, the only confirmed source of CP violation is the Kobayashi-Maskawa (KM) phase present in the CKM matrix \cite{cabibbo}\cite{cpvsm}, which arises when we move the quarks from flavour to mass eigenstate in the Standard Model (SM)}. However, we expect to find more sources of CP violation owing to the observed matter-antimatter asymmetry in the universe \cite{asymmetrycond}. Thus, it is imperative to look for CP-violating observables, especially those which are very small or zero in SM, because if they deviate even slightly from zero (which can be checked by a null test), it would not just be a discovery of a new source of CP violation, \cng{but} also be a smoking-gun signal of new physics (NP). 
\par

In this article, we study the $B_s^0 \rightarrow \phi \phi$ decay (where $\phi(1020)$ is implied throughout this paper), which is a \cng{$B \rightarrow VV$ type} pure penguin process. \cng{$B \rightarrow VV$ type processes have been extensively studied in the literature }\cite{bvv1,bvv2,bvv11,bvv12,bvv13,bvv14,bvv15,bvv16,bvv17,bvv18,bvv19,bvv20,qcd1np,qcd2np,qcd3np}. The presence of penguin quantum loop makes it an excellent probe to search for new heavy particles and being a purely penguin-type decay keeps it free from tree-penguin interference contamination, making it a clean observable to search for NP \cite{qcd1np}\cite{qcd2np}\cite{qcd3np}. The object of interest is going to be the phase in the interference of the direct decay of $B^0_s$ mesons and decay via mixing of $B_s^0 - \bar{B}_s^0$ to CP eigenstates, which is a CP-violating parameter. This phase is expected to be very small in SM ($-2\beta_s \approx O(\lambda^2)$). In this paper, we will be presenting a new scheme for the interference phases of different helicities within the framework of our chosen model of study\cng{, which is constructed by adding the Chromomagnetic dipole operator (and its chirally-flipped counterpart) to our Hamiltonian.}
\par
The objective of this article is threefold. The first is to show the power of angular decay distribution: it can help segregate the final state when it is a mixture of different helicities. Combining it with a $ P \rightarrow VV$ type decay ($P$-pseudoscalar particle and $V$-vector particle) gives us access to three (helicity) amplitudes instead of one, meaning we can go beyond the assumption of helicity-independent phases to probe three CP-violating phases, and possibly three new indicators of NP. 
\par
Secondly, we present a new scheme for the interference phases, based on the hierarchy of helicity amplitudes arising in our model, which is different from the ones LHCb used in its fits \cite{lhcb}. We also note the fact that LHCb's objective is to do a null test of the {interference phase}, without any regard to its origin (decay or mixing). However, we specifically assume that the weak phase is coming from decay amplitude, not mixing amplitude. {Consequently, we change the form of helicity/transversity amplitude to include a CP-violating decay phase.} This modifies the coefficients of time-dependent part of amplitude, which we present in Table~\ref{timedeptablestrongCP}. 
In addition, we investigate the $B^0_d \rightarrow \phi K_s$ decay amplitude with our NP Hamiltonian.  We show that the Belle(II)'s $B^0_d \rightarrow \phi K_s$ decay measurement  along with the LHCb's $B^0_s \rightarrow \phi\phi$ measurement can provide the chirality of NP in our model, as long as the signs of cosine of strong phases of these decays can be obtained from the theory.
\par
Lastly, we perform a sensitivity study to illustrate the new scheme of experimental analysis we are proposing. We perform a fit with two pseudo datasets (based on two sets of result of LHCb) to calculate the sensitivities of the CP-violating parameters, which also act as null test parameters for new physics.
\par
\mycomment{The model we choose to use in our study is that of the \emph{Chromomagnetic dipole operator} \emph{$O_{8g}$}, which, for $\bar{b}\rightarrow \bar{s}g$ process, is given as follows:
\begin{align}\label{chromomagneticoperator}
O_{8g} = \frac{g_s}{8\pi^2} m_b \bar{b}_{\alpha} \sigma^{\mu\nu} (1 + \gamma^5) \frac{\lambda^a_{\alpha\beta}}{2}s_{\beta}G^a_{\mu\nu}.
\end{align}
To keep it very general, we work with both left- and right-handed {\it New Physics} currents; this is done by adding the Chromomagnetic operator with opposite quark chiralities (i.e. by flipping sign of $\gamma^5$) (see Eq.~\eqref{effhamiltonian}).
Our choice of model is motivated by the fact that the Chromomagnetic operator has been observed to be sensitive to contributions from various NP models \cite{pnp1}\cite{pnp2}\cite{pnp3}\cite{pnp4}\cite{pnp5}\cite{pnp6}\cite{pnp7}\cite{pnp8}. There are some NP models that give the same contribution to both decay and mixing amplitudes. This causes the contribution to cancel out in the interference phase, and they remain undetectable in $B^0_s \rightarrow \phi\phi$ decay. However, since Chromomagnetic operator only contributes to the decay amplitudes, a NP contribution manifesting itself through this operator can be very well detected via this channel (the $B_s^0 - \bar{B}_s^0$ mixing amplitude has already been well constrained by previous measurements 
\cite{cpvinosc}\cite{lhcbjpsimixing}, so we do not focus on it in this work). We emphasise that the choice of the model is not unique, as other models could possibly give similar contributions. Our objective is simply to show how to fully utilise the power of angular distribution when coupled with $P \rightarrow VV$ decays, and provide a new fit scenario, which can be used to check for the presence of NP.
}
\par 
\cng{The organisation of the article is as follows: In Section~\ref{sec:angdecaydist}, we describe the angular decay distribution of the $B^0_s \rightarrow \phi(\rightarrow K^+K^-)\phi(\rightarrow K^+K^-)$. In Section~\ref{sec:npsearchviacp}, we talk about the CP-violating parameters in SM and in the presence of a NP amplitude. In Section~\ref{sec:npmodel}, we introduce our NP Hamiltonian and do a helicity/transversity analysis in order to pinpoint the effect of NP in the correct transversity amplitude, based on which we present our new phase scheme in Section~\ref{sec:newphasescheme}. Following this phase scheme, we do a sensitivity study on the CP-violating parameters with two pseudo datasets in Section~\ref{sec:sensitivitystudy}. Finally, we show that under certain conditions, the results of $B_d^0 \rightarrow \phi K_s$ from Belle(II) can be used to complement the results of $B_s^0 \rightarrow \phi\phi$ to find the chirality of NP. }

\section{Angular decay distribution}\label{sec:angdecaydist}
\begin{figure}[t]
\setlength{\unitlength}{1mm}
  \centering
  \begin{picture}(140,60)
    \put(0,-1){
      \includegraphics*[width=140mm]{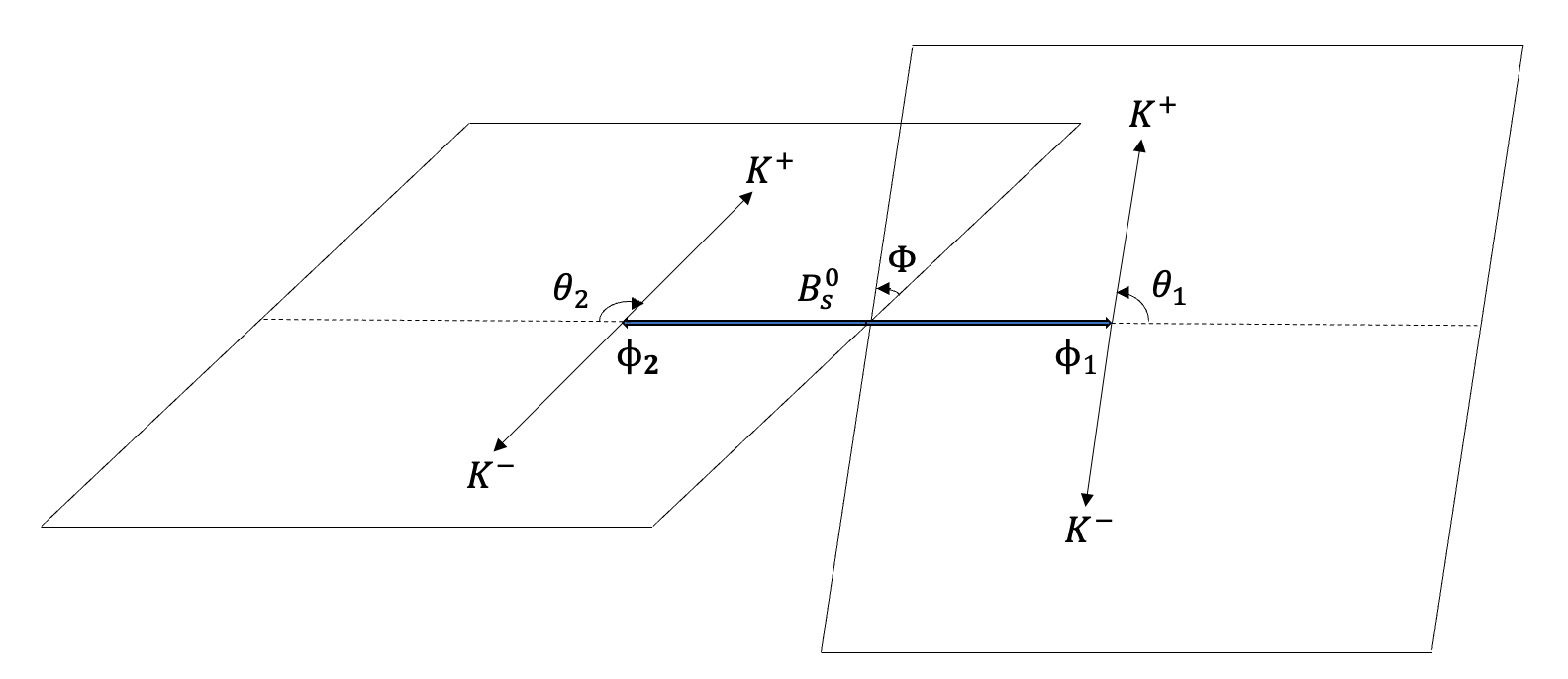}
    }
  \end{picture}
  \caption{\small Decay angles for the \cng{$B_s^0 \rightarrow \phi(\rightarrow K^+K^-) \phi(\rightarrow K^+K^-)$} decay, where  
$\theta_{1(2)}$ is the angle between the $K^+$ momentum in the $\phi_{1(2)}$ meson rest frame and the $\phi_{1(2)}$ momentum in the $B_s^0$ rest frame. $\Phi$ is the angle between the two $\phi$ meson decay planes. \cng{The angular conventions used are defined in detail in Appendix~\ref{angularconventions}.}}
\label{fig:decayangles}
\end{figure}

The angular decay distribution for \cng{$B^0_s \rightarrow \phi(\rightarrow K^+K^-)\phi(\rightarrow K^+K^-)$} decay can be described by the help of three angles as shown in Figure \ref{fig:decayangles}. A random choice is made for which $\phi$ meson is used to determine $\theta_1$ and $\theta_2$. The power of angular analysis is that it can disentangle the final states of $B^0_s \rightarrow \phi\phi$ decay (which is a mixture of CP eigenstates) and we get access to three (helicity/transversity) amplitudes instead of one, meaning we can probe three CP-violating phases, and possibly three new indicators of NP. We will neglect the contribution of scalar $f_0(980)$ resonance, as it can be removed by appropriate experimental cuts \cite{lhcb}\cite{hflav}.
\cng{The amplitude then for this process is given by}
\begin{align}\label{amplitude}
\mathcal{A}(t,\theta_1,\theta_2,\Phi) &= A_0(t)\cos\theta_1\cos\theta_2 
+\frac{A_\parallel(t)}{\sqrt{2}}\sin\theta_1\sin\theta_2\cos\Phi \nonumber \\
&+i\frac{A_\perp(t)}{\sqrt{2}}\sin\theta_1\sin\theta_2\sin\Phi,
\end{align}
where $A_0$ is the longitudinal CP-even, $A_{\parallel}$ is the transverse-parallel CP-even and $A_{\perp}$ is the transverse-perpendicular CP-odd transversity amplitude. The resulting angular decay distribution is proportional to square of the amplitude in Eq.~\eqref{amplitude} and has 6 terms \cite{transversity}:
\begin{align}\label{angdist}
\frac{d^4\Gamma}{d t d \cos\theta_1\,d \cos\theta_2 \,d\Phi} \propto |\mathcal{A}(t,\theta_1,\theta_2,\Phi)|^2
= \frac{1}{4}\sum_{i=1}^{6} K_i (t) f_i (\theta_1,\theta_2,\Phi)
\end{align}
The angular dependence contained in $f_i(\theta_1,\theta_2,\Phi)$ is as follows:
\begin{align}
|\mathcal{A}(t,\theta_1,\theta_2,\Phi)|^2
&= \frac{1}{4} \bigl[4K_1(t) \cos^2\theta_1 \cos^2\theta_2 + K_2(t) \sin^2\theta_1 \sin^2\theta_2 (1+\cos2\Phi) \nonumber \\
&+ K_3(t) \sin^2\theta_1 \sin^2\theta_2 (1-\cos2\Phi) -2K_4(t) \sin^2\theta_1 \sin^2\theta_2 \sin2\Phi \nonumber \\
&+ \sqrt{2}K_5(t)  \sin2\theta_1 \sin2\theta_2 \cos\Phi - \sqrt{2}K_6(t) \sin2\theta_1 \sin2\theta_2 \sin\Phi \bigr].
\end{align}
The time dependence is contained in $K_i(t)$ which is defined as
\begin{align} \label{timedep}
\begin{aligned}
K_i(t)=N_i e^{-\Gamma_s t} \bigg[ 
&a_i \cosh\left(\frac{1}{2}\Delta \Gamma_s t\right) + b_i \sinh\left(\frac{1}{2}\Delta\Gamma_s t\right) + 
c_i \cos(\Delta m_s t) + d_i\sin(\Delta m_s t) \bigg].
\end{aligned}
\end{align}
{The coefficients $a_i,b_i,c_i$ and $d_i$ are the LHCb experimental observables given in  Table~\ref{timedeptablestrongCP}.} \cng{The structure of these coefficients depend on the form of amplitudes $A_{0,\parallel,\perp}(t)$, defined in Section~\ref{cpvnpparam}}. $\Delta \Gamma_s \equiv \Gamma_L - \Gamma_H$ is decay-width difference between the light and heavy $B_s^0$ mass eigenstate, $\Gamma_s \equiv (\Gamma_L + \Gamma_H)/2$ is the average decay width and $\Delta m_s \equiv m_H - m_L$ is the mass difference between the heavy and light $B_s^0$ mass eigenstate, and also the $B_s^0 - \bar{B}_s^0$ oscillation frequency. Their values are $\Delta \Gamma_s = 0.086 \pm 0.006 \text{ ps}^{-1}$ and $\Gamma_s = 0.6646 \pm 0.0020 \text{ ps}^{-1}$ \cite{hflav}, and the oscillation frequency is constrained by the LHCb measurement to be $\Delta m_s = 17.768 \pm 0.023 \text{ (stat)} \pm 0.006 \text{ (syst)} \text{ ps}^{-1}$ \cite{oscfreq}. 
\section{Search for new physics \cng{via CP observables}}\label{sec:npsearchviacp}
\subsection{CP-violating \cng{quantities} in the Standard Model}\label{cpvsm}
Before looking at how to search for NP, we must know the SM predictions \cite{angdist}. The phase in the interference of decay with and without mixing is almost zero in SM in $B^0_s \rightarrow \phi\phi$ decays because the KM phase in $B_s^0$ decay amplitude cancels the one arising from the $B_s^0 - \bar{B}_s^0$ mixing box diagram (considering the dominant $t$-quark contribution). But for a more accurate prediction of phase (to higher orders in $\lambda$), we need to consider the contribution of $u$ and $c$-quarks too. 
These contributions can arise due to QCD rescattering $c\bar{c} \rightarrow q\bar{q}$ and $u\bar{u} \rightarrow q\bar{q}$ ($q = d,s$) from tree operators $\bar{b} \rightarrow \bar{c}c\bar{s}$ and  $\bar{b} \rightarrow \bar{u}u\bar{s}$, respectively, and may have a contribution up to around $20-30\%$  \cite{rescattering} of the dominant top amplitude.
Taking into account these contributions, the SM amplitude for $\bar{b} \rightarrow \bar{s}$ decay for a given helicity '$k$' can be written as
\begin{align}
A_k^{\rm{SM}} = \lambda_t P_{t,k} + \lambda_c R_{c,k} + \lambda_u R_{u,k}
\end{align}
$\lambda_q = V_{qb}^* V_{qs}$ is the CKM matrix element.
Here, while $P_{t,k}$ arises due to gluonic penguin with a $\bar{t}$-quark in the loop, $R_{c,k}$ and $R_{u,k}$ are the rescattering contribution. Using unitarity of CKM matrix to eliminate the $c$-quark contribution and writing strong phases explicitly, we get 
\begin{align}\label{ampSM}
\begin{aligned}
A_k^{\rm{SM}} &= |V_{tb}^* V_{ts}| e^{-i\beta_s} |PR_{tc,k}| e^{i\delta_{tc,k}} + |V_{ub}^* V_{us}|e^{i\gamma} |RR_{uc,k}|e^{i\delta_{uc,k}} \\
&=  |V_{tb}^* V_{ts}| e^{-i\beta_s} |PR_{tc,k}| e^{i\delta_{tc,k}}  \left[ 1 + r_k^{\rm{SM}} e^{i(\gamma + \beta_s)} e^{i(\delta_{uc,k} - \delta_{tc,k})} \right] \\
&= |A_k^{\rm{SM}}| e^{i\phi^{\rm{SM}}} e^{i\delta_k^{\rm{SM}}},
\end{aligned}
\end{align}
where $PR_{tc,k} = P_{t,k} - R_{c,k}$, $RR_{uc,k} = R_{u,k} - R_{c,k}$, $\delta$ denote the {SM} strong phases, $\beta_s = arg(\frac{-V_{ts}V_{tb}^*}{V_{cs}V_{cb}^*}) \approx \eta \lambda^2$ and $r_k^{\rm{SM}} = \frac{|V_{ub}^* V_{us}| |RR_{uc,k}|}{|V_{tb}^* V_{ts}|  |PR_{tc,k}|}$. Assuming that that the rescattering contribution is around $20-30\%$ of the dominant penguin amplitude, we can write $\frac{|RR_{uc,k}|}{|PR_{tc,k}|} = O(\lambda)$, as $\lambda \approx 0.22$. Therefore, we have $r_k^{\rm{SM}} = O(\lambda^3)$. This would make all the CP-violating observables (like indirect CP asymmetry, triple product asymmetries etc.) $O(\lambda^3)$ (or smaller). This can be inferred from the fact that CP violation occurs when two amplitudes interfere. Since one of the amplitudes is much smaller than the other (by O($\lambda^3$)), the observed CP-violating observables are $O(\lambda^3)$ too. This becomes one of the key point in our NP search: since we know the order of CP-violating observables in SM, any observable that is larger than $O(\lambda^3)$ would be a clear signal of NP.
\par
\cng{We emphasise that we only search for NP via the CP-violating observables, namely, the direct CP violation parameter and the interference phase, since their SM values are known with much better precision (as compared to other observables like branching ratios). Having a better handle on theoretical predictions allow us to look for small NP effects - especially in the case of $B_s \rightarrow \phi \phi$, where we basically have to do a null-test on the CP observables.}
\subsection{CP-violating \cng{quantities in the presence of new physics}: \\Parametrisation} \label{cpvnpparam}
In this study, as mentioned before, we are only probing CP-violating phases in the decay; thus, our parametrisation is done accordingly. Here, for generality, we include both left- and right-handed currents (which could arise from several NP models), which could give rise to new CP-violating phase(s). Also, we assume $|\frac{q}{p}| = 1$ \cite{mixingcpasymmetry}. 
\par
The helicity/transversity amplitudes, with helicity/transversity '$k$' are written as \cite{rosner} 
\begin{align}
\begin{aligned}
A_k(t) &= \langle (\phi\phi)_k | \Ham_{\rm{eff}} | B_s^0(t) \rangle = g_+(t) A_k + \frac{q}{p}g_-(t)\bar{A}_k  \\
\bar{A}_k(t) &= \langle (\phi\phi)_k | \Ham_{\rm{eff}} | \bar{B}_s^0(t) \rangle = g_+(t) \bar{A}_k + \frac{p}{q} g_-(t) A_k.
\end{aligned}
\end{align}
where $g_+(t)$ and $g_-(t)$ describe the time evolution of $B^0_s$ and $\bar{B}^0_s$, respectively.
Using Eq.~\eqref{ampSM} and adding a NP component, the amplitude at $t=0$ can be written as:
\begin{align} \label{ampNP}
\begin{aligned}
A_k(0) \equiv A_k &= A_k^{\rm SM}+A_k^{\rm NP} \\
&= |A_k^{\rm SM}| e^{i\delta_k^{\rm SM}}e^{i\phi^{\rm SM}} + |A_k^{\rm NP}| e^{i\delta_k^{\rm NP}}e^{i\phi_k^{\rm NP}}  \\
&= |A_k^{\rm SM}| e^{i\delta_k^{\rm SM}}e^{i\phi^{\rm SM}}  \left( 1 + r_k^{\rm NP} e^{i(\phi_k^{\rm NP} - \phi^{\rm SM})} e^{i(\delta_k^{\rm NP} - \delta_k^{\rm SM})} \right) \\
&= |A_k^{\rm SM}| e^{i\delta_k^{\rm SM}}e^{i\phi^{\rm SM}} X_k e^{i\theta_k}, 
\end{aligned}
\end{align}
where in the last line, we denote the quantity in the parenthesis as $X_k e^{i\theta_k}$ and $r_k^{\rm NP} = \frac{|A_k^{\rm NP}|}{|A_k^{\rm SM}|}$. The phase $\theta_k$ is a \emph{mixture of weak and strong phases}. \footnote{Notice that if we assume $\delta_k^{\rm SM} = \delta_k^{\rm NP}$, then the phase $\theta_k$ would be a purely weak phase. In such a case the interference phase in Eq.~\eqref{intphase} would not just tell us about the presence of NP, it would also tell us the value of CP-violating (weak) phase in the decay amplitude. However, we work in the most general case in this calculation, as we only wish to probe for NP.}Similarly for the CP-conjugate amplitude, the expression is ($\eta_k$ is the CP eigenvalue of the transversity state, with $\eta_{\perp} =-1$ and $\eta_{0,\parallel} =1$)
\begin{align}\label{ampNPconj}
\begin{aligned}
\bar{A}_k &= \eta_k |A_k^{\rm SM}| e^{i\delta_k^{\rm SM}}e^{-i\phi^{\rm SM}}  \left( 1 + r_k^{\rm NP} e^{-i(\phi_k^{\rm NP} - \phi^{\rm SM})} e^{i(\delta_k^{\rm NP} - \delta_k^{\rm SM})} \right) \\
&= \eta_k |A_k^{\rm SM}| e^{i\delta_k^{\rm SM}}e^{-i\phi^{\rm SM}} X_k^c e^{i\theta_k^c}.
\end{aligned}
\end{align}
Recalling that $\arg(q/p) = 2\beta_s \approx 2\phi^{\rm SM}$, we finally get 
\begin{align}\label{intphase}
\frac{q}{p}\frac{\bar{A}_k}{A_k} = \eta_k \lambda_k e^{-i(\theta_k - \theta_k^c)},
\end{align}
where $\lambda_k \equiv \frac{|\bar{A}_k|}{|A_k|} = \frac{X^c_k}{X_k}$ becomes the direct CP violation measurement parameter: $\lambda_k \neq 1$ implies direct CP violation is present in the decay. Since in SM, $\lambda_k =1$ for all helicities, the deviation of this value from $1$ (by more than $O(\lambda^3)$) would be a clear signal for NP, i.e. $\lambda_k-1$ is a null-test parameter for NP.  Another quantity that can be used for NP search is the interference phase $\theta_k-\theta_k^c$. In SM, this quantity is zero, as explained in Section~\ref{cpvsm}. Therefore, the deviation of this quantity from zero (by more than $O(\lambda^3)$) would be a signal of NP, i.e. $\theta_k-\theta_k^c$ is also a null-test parameter for NP.   One must note that there is one special case when neither of these two parameters would be able to detect the presence of NP: it is the case when $\phi^{\rm NP}_k = \phi^{\rm SM}$. In this case, $\lambda_k = 1$ and $\theta_k - \theta_k^c =0$ and NP cannot be detected by CP-violating observables.
\par
\cng{Here, we take a moment to explain the $\eta_k$ factors used. When we write the CP conjugate decay, we replace the particles by their antiparticles. The effect of this replacement on the helicity angle is $\phi \rightarrow 2\pi - \phi$, which gives rise to a negative sign in those terms which contain amplitudes having a negative CP parity ($A_\perp$ in our case). Therefore, using $\eta_k$ in the definition of amplitude allows us to use the same angular functions for $B^0_s$ and $\bar{B}^0_s$ decays, which facilitates calculations in untagged samples \cite{angdist}.}
\par
The time-dependent amplitude is given by
\begin{align}\label{timedepamp}
\begin{aligned}
A_k(t) &= A_k \left[ g_+(t) + g_-(t) \frac{q}{p} \frac{\bar{A}_k}{A_k} \right] \\
A_k(t) &= |A_k^{\rm SM}| X_k e^{i\delta_k^{\rm SM}}e^{i\phi^{\rm SM}} e^{i\theta_k} \left[ g_+(t) + g_-(t) \eta_k \lambda_k e^{-i(\theta_k - \theta_k^c)} \right]. 
\end{aligned}
\end{align} 
The coefficients of the time-dependent terms in Eq.~\eqref{timedep}, obtained by using Eq.~\eqref{timedepamp},  are given in Table~\ref{timedeptablestrongCP}.
\cng{Our time-dependent amplitude differs from the one given by LHCb, because while LHCb has used the amplitude $A_k = |A_k|e^{i\delta_k}$, our amplitude contains both SM and NP contribution, and both contain strong and weak phases (see Eq.~\eqref{ampNP}).\footnote{The mixing-induced CP violation (i.e. the CP violation in interference of decay with and without mixing) cannot tell us if the original source of this effect is coming from the dynamics of $\Delta B=2$ (mixing) sector or that of $\Delta B=1$ (decay) sector, as long as we are working with only one final state (or one pair of CP-conjugate final state). We need information from at least one more final state to decide unambiguously the presence of direct CP violation and/or CP violation in mixing \cite{bigisandracpv}.} Thus, we have the phase $\phi^{\rm{SM}}+\theta_k$ along with $\delta_k^{\rm SM}$ (contrary to LHCb equation where there only is $\delta_k^{\rm SM}$ outside the bracket). In addition, because we have both SM and NP amplitudes, we get two different mixed phases ($\theta_k$ and $\theta_k^c$) coming from $A_k$ and $\bar{A_k}$, and thus the interference phase is $\theta_k-\theta_k^c$, contrary to LHCb's equation, where the interference phase is simply $\phi_{s,k}$.\footnote{\cng{This increase in number of parameters will make the search more sensitive to NP. However, this comes at a price: it becomes more difficult to make the fit converge. Thus, we need model-dependent simplifying assumptions to reduce the free parameters, as we'll show in the subsequent sections, to make the fit converge.}}} This changes the coefficients in Table~\ref{timedeptablestrongCP} with respect to the ones given by LHCb \cite{lhcb}.\footnote{\cng {If $\theta_k$ and $\theta^c_k$ are helicity independent, they will cancel out when we write terms of type $A_i(t)A_k^*(t)$ or $|A_i(t)|^2$.} Then our formula and that of LHCb would exactly be the same.} For simplicity of notation, we simply denote $\delta_k^{\rm SM}$ as $\delta_k$ in the rest of the paper.

\section{\cng{New physics model}}\label{sec:npmodel}
\cng{The model we choose to use in our study is that of the \emph{Chromomagnetic dipole operator} \emph{$O_{8g}$}, which, for $\bar{b}\rightarrow \bar{s}g$ process, is given as follows:
\begin{align}\label{chromomagneticoperator}
O_{8g} = \frac{g_s}{8\pi^2} m_b \bar{b}_{\alpha} \sigma^{\mu\nu} (1 + \gamma^5) \frac{\lambda^a_{\alpha\beta}}{2}s_{\beta}G^a_{\mu\nu}.
\end{align}
Though Chromomagnetic operator is a SM operator, it is suppressed by $b$-quark mass $m_b$ (and its chirally-flipped counterpart is suppressed by $s$-quark mass $m_s$). However, it is very sensitive to several NP models, like the Left-Right symmetric class of models or SUSY, where it can undergo \emph{chiral enhancement} to overcome the quark mass suppression \cite{pnp1,pnp2,pnp3,pnp4,pnp5,pnp6,pnp7,pnp8}. In addition, there are some NP models that give the same contribution to both decay and mixing amplitudes. This causes the contribution to cancel out in the interference phase, and they remain undetectable in $B^0_s \rightarrow \phi\phi$ decay. However, since Chromomagnetic operator only contributes to the decay amplitudes, a NP contribution manifesting itself through this operator can be very well detected via this channel (the $B_s^0 - \bar{B}_s^0$ mixing amplitude has already been well constrained by previous measurements 
\cite{cpvinosc}\cite{lhcbjpsimixing}, so we do not focus on it in this work).}

\par
\mycomment{In this section, we apply the technique of helicity amplitude analysis to a model with operator $O_{8g}$ with both left- and right-handed currents. 
As mentioned before, our choice of operator $O_{8g}$ is motivated by the fact that it is sensitive to NP, and that it contributes only to $B_s^0$ decay amplitude, and not $B_s^0 - \bar{B}_s^0$ mixing amplitude.
\par}
Starting from the effective Hamiltonian for $\Delta B = 1$ decay, it is given by $(q \in \{d,s\})$
\begin{align}\label{effhamiltonian}
\Ham_{\rm{eff}} =-\frac{G_F} {\sqrt{2}} V^*_{tb}V_{tq}  
 \left[ \sum_{i=3}^{6} (C_i^{\rm{SM}} O_i) + C_{8g} O_{8g} + \tilde{C}_{8g} \tilde{O}_{8g}  \right] + \rm{h.c.}
\end{align}
The operators are given by $(q' \in \{u,d,s,c\})$
\cng{
\begin{align}
\begin{aligned}
O_3 &= (\bar{b}_{\alpha} q_{\alpha})_{V-A} \sum_{q'} (\bar{q}'_{\beta} q'_{\beta})_{V-A},  \quad \quad
O_4 &= (\bar{b}_{\beta} q_{\alpha})_{V-A} \sum_{q'} (\bar{q}'_{\alpha} q'_{\beta})_{V-A},  \\
O_5 &= (\bar{b}_{\alpha} q_{\alpha})_{V-A} \sum_{q'} (\bar{q}'_{\beta} q'_{\beta})_{V+A},  \quad\quad
O_6 &= (\bar{b}_{\beta} q_{\alpha})_{V-A} \sum_{q'} (\bar{q}'_{\alpha} q'_{\beta})_{V+A} ,
\end{aligned}
\end{align}
}
with the notation $(\bar{a} b) _{V \pm A}(\bar{c} d) _{V \pm A} = \bar{a} \gamma^{\mu}(1 \pm \gamma^5)b \;  \bar{c} \gamma_{\mu}(1 \pm \gamma^5)d $.
\par
Here, we only include the gluonic penguin operators. \cng{The operator with tilde is obtained by changing the sign of $\gamma_5$ term in the definition of $O_{8g}$ to obtain the chirally-flipped counterpart. }
\mycomment{
We can express $\ptwiddle{O}_{8g}$, given in Eq.~\eqref{chromomagneticoperator}, in terms of operators $\ptwiddle{O}_{3-6}$, the details of which are given in Appendix~\ref{cg}. Once we do this, we can write the effective Hamiltonian as
\begin{align}
\Ham_{\rm{eff}} =-\frac{G_F} {\sqrt{2}} V^*_{tb}V_{tq}  
 \sum_{i=3}^{6} (C_i^{\rm{SM}} O_i  + C_i^{\rm{L}} O_i + C_i^{\rm{R}} \tilde{O}_i) .
\end{align}
\par
Once we have the Hamiltonian, we can write the amplitude for $B_s^0 \rightarrow \phi \phi_s$ \footnote{We have marked one of the $\phi$ with a subscript 's', just to differentiate between two mesons for convenience in theoretical calculations (s denoting that it contains the spectator quark). Experiments cannot distinguish between the two.} for SM, and left- and right-handed NP currents. In the Naive Factorisation method, they are given by 
\begin{align}\label{phiphiamp}
\begin{aligned}
\Amp^{\rm{SM}}_{\phi \phi} &= - \frac{G_F}{\sqrt{2}} V^*_{tb} V_{ts} \xi^{\rm{SM}}_{345}
 \langle \phi | \bar{s} \gamma^{\mu}(1 - \gamma^5) s | 0 \rangle \langle \phi_s | \bar{b} \gamma_{\mu}(1 - \gamma^5) s | B_s^0 \rangle \\
\Amp^{\rm{L}}_{\phi \phi} &= - \frac{G_F}{\sqrt{2}} V^*_{tb} V_{ts} \xi^{\rm{L}}_{345}
 \langle \phi | \bar{s} \gamma^{\mu}(1 - \gamma^5) s | 0 \rangle \langle \phi_s | \bar{b} \gamma_{\mu}(1 - \gamma^5) s | B_s^0 \rangle \\
\Amp^{\rm{R}}_{\phi \phi} &= - \frac{G_F}{\sqrt{2}} V^*_{tb} V_{ts} \xi^{\rm{R}}_{345}
 \langle \phi | \bar{s} \gamma^{\mu}(1 + \gamma^5) s | 0 \rangle \langle \phi_s | \bar{b} \gamma_{\mu}(1 + \gamma^5) s | B_s^0 \rangle
\end{aligned}
\end{align}
with $ \xi_{345}^{p} = \xi_3^p + \xi_4^p + \xi_5^p \quad (p \in \{\rm{SM},\rm{L},\rm{R}\})$ are the combinations of the Wilson coefficients given by \cite{vsa}
\begin{align}
\xi_i^p =C^{p}_i + \frac{1}{N_c}C^{p}_{i+1}  \quad \text{(i is odd)} \quad\quad \xi_i^p =C^{p}_{i+1} + \frac{1}{N_c}C^{p}_{i} \quad \text{(i is even)},
\end{align}
where $N_c$ are the number of colours. The Wilson coefficients and related details of calculations are given in Appendix~\ref{cg}.
\cng{The matrix elements for the left- and right-handed currents are now given by
\begin{align}
\begin{aligned}
 &\langle \phi | \bar{s} \gamma^{\mu}(1 \mp \gamma^5) s | 0 \rangle \langle \phi_s | \bar{s} \gamma_{\mu}(1 \mp \gamma^5) b | B \rangle = f_{\phi} m_{\phi} \\
 & \Big[ \mp (m_B + m_{\phi})(\epsilon_{\phi_S}^* . \epsilon_{\phi}^*) A_1(q^2)  \pm
  (\epsilon_{\phi_s}^* . p_{\phi})(\epsilon_{\phi}^* .  p_{\phi_s}) \frac{2 A_2(q^2)}{m_B + m_{\phi}}\\
 &+i \epsilon_{\mu\nu\alpha\beta} \epsilon_{\phi}^{*\nu}c\epsilon_{\phi_s}^{*\nu}p_{\phi}^{\alpha} p_{\phi_s}^{\beta} \frac{2V(q^2)}{m_B + m_{\phi}}  \Big].
\end{aligned}
\end{align}
\emph{The $A_{1,2} (q^2)$ terms change sign due to $V \pm A$ structure of current} - this would be a key point later in the discussion.
}
}
\par
Once we have the model clearly defined with Hamiltonian and amplitudes, we can move on to the helicity analysis. As mentioned before, the advantage in $P\rightarrow VV$ type decays is that the final state can be split into three helicity states, which gives us access to three amplitudes, whose sum makes up the total amplitude. 
The general form of helicity amplitude for the process $B(M,p) \rightarrow V_1(\epsilon_1,M_1,k_1) + V_2(\epsilon_2,M_2,k_2)$ is given by \footnote{We use the convention $\epsilon^{0123} = 1$. The opposite convention would simply interchange the definition of $H_+$ and $H_-$, without affecting $H_0$} \cite{helampkramer}
\begin{align}\label{helampgen}
\begin{aligned}
H_{\lambda} &= a (\epsilon^*_1(\lambda).\epsilon^*_2(\lambda)) + \frac{b}{M_1 M_2} (\epsilon^*_1(\lambda) . k_2) (\epsilon^*_2 (\lambda).  k_1) \\
&+ \frac{i c}{M_1 M_2} \epsilon_{\mu \nu \rho \sigma} \epsilon_1^{*\mu} (\lambda) \epsilon_2^{* \nu}(\lambda) k_1^{\rho} k_2^{\sigma},
\end{aligned}
\end{align}
where $\lambda = \{+,-,0\}$ is the polarisation of final state.  $a,b$ and $c$ are the \emph{Invariant Amplitudes}. Putting the polarisation vectors for the three different polarisations, we get
\begin{align} \label{helamp}
\begin{aligned}
H_0 = ax + b(x^2 -1) \quad \quad \quad H_{\pm} = a \pm c \sqrt{x^2 -1},
\end{aligned}
\end{align}
where $x =  \frac{M_B^2 - M_1^2 - M_2^2}{2 M_1 M_2}$.
A more convenient basis to work than helicity basis is \emph{transversity basis} \cite{transversity}. To go there, we first note that by angular momentum conservation, the final state helicities could only be $|++ \rangle , |-- \rangle $ and $|00 \rangle$.  Since $P|++ \rangle =|-- \rangle$ and $P|00 \rangle = |00 \rangle$, we can define parity eigenstates with eigenvalues $\pm 1$ as \cite{bigisandracpv}
\begin{align}
 | \parallel \,\rangle = \frac{1}{\sqrt{2}} (|++ \rangle + |-- \rangle) \quad\quad\quad | \perp \,\rangle = \frac{1}{\sqrt{2}} (|++ \rangle - |-- \rangle).
\end{align}
\cng{These are called} \emph{transversity amplitudes}, denoted by $A_{k}$, with $k=\{0,\parallel,\perp\}$, as given in Eq.~\eqref{amplitude}.
Clearly 
\begin{align} 
\Ampl_{\parallel , \perp} = \frac{1}{\sqrt{2}} ( H_+ \pm H_- ) \quad\quad\quad \Ampl_0 = H_0.
\end{align}
Thus, 
\begin{align}\label{transamp}
A_0 = ax + b(x^2 -1), \quad\quad A_{\parallel} = \sqrt{2}a, \quad\quad
A_{\perp} = \sqrt{2(x^2 -1)}c.
\end{align}
\mycomment{
Now we can compare the general helicity amplitude structure in Eq.~\eqref{helampgen} with our $B_s^0 \rightarrow \phi\phi$ amplitudes given in Eq.~\eqref{phiphiamp} to get the invariant amplitudes. \cng{On comparing, we can see that the invariant amplitudes $a$ and $b$ have opposite signs for left- and right-handed amplitudes. Using these invariant amplitudes, we can write the helicity and transversity amplitudes, defined in Eq.~\eqref{helamp} and Eq.~\eqref{transamp}, respectively. \emph{Since $A_0$ and $A_{\parallel}$ are functions of $a$ and $b$, they will also have opposite signs for left- and right-handed amplitudes}.}
\par
}

\cng{Let us see the hierarchy of amplitudes predicted by the $V-A$ structure of current. The hierarchy is $H_0 > H_{+} > H_{-}$ (interchange $+$ and $-$ signs for $\bar{B}^0_s$ decay) \cite{kagan}, with the approximate ratio $H_0 : H_+ : H_- \sim 1: (\frac{m_{\phi}}{m_B}) :(\frac{m_{\phi}}{m_B})^2 $ \cite{helicityhierarchy1}\cite{kagan}. However, it is well known that this hierarchy is not observed experimentally in $B \rightarrow VV$ decays. A large transverse polarisation was first observed in $B_d\rightarrow \phi K^*$ \cite{phikstar} (and then later in $B_d \rightarrow J/\psi\phi$ \cite{lhcbjpsimixing}, $B_s \rightarrow \phi \phi$ \cite{lhcb} etc.)  which gave rise to intense theoretical and experimental studies of charmless $B \rightarrow VV$ decays. Several theoretical papers have been written to go beyond the naive factorisation method and use more sophisticated tools (like QCDf, pQCD, SCET etc.) to compute these decays more accurately \cite{qcd1np}\cite{qcd2np}\cite{qcd3np}\cite{kagan}. It has been pointed out in \cite{qcd2np}\cite{kagan}\cite{annihilationdiagram} that a major contributor to transverse amplitudes are the \emph{annihilation diagrams} which can explain the large fraction of transverse amplitudes observed experimentally.  }

\mycomment{Let us see the hierarchy of amplitudes predicted by the $V-A$ structure of current. For the case of $B^0_s$ decay, consider the current $(\bar{b}s)_{V-A}$: The longitudinal amplitude $H_0$ is the dominant one. There is a spin-flip suppression in $H_{+}$ amplitude, and the $H_{-}$ amplitude has a chirality suppression in the light s-quark along with spin-flip suppression \cite{helicityhierarchy1}\cite{helicityhierarchy2}. Therefore, the expected hierarchy is $H_0 > H_{+} > H_{-}$ (interchange $+$ and $-$ signs for $\bar{B}^0_s$ decay) \cite{kagan}. The approximate ratio is $H_0 : H_+ : H_- \sim 1: (\frac{m_{\phi}}{m_B}) :(\frac{m_{\phi}}{m_B})^2 $ \cite{helicityhierarchy1}\cite{kagan}. However, it is well known that this hierarchy is not observed experimentally in $B \rightarrow VV$ decays. A large transverse polarisation was first observed in $B_d\rightarrow \phi K^*$ \cite{phikstar} (and then later in $B_d \rightarrow J/\psi\phi$ \cite{lhcbjpsimixing}, $B_s \rightarrow \phi \phi$ \cite{lhcb} etc.)  which gave rise to intense theoretical and experimental studies of charmless $B \rightarrow VV$ decays. Several theoretical papers have been written to go beyond the naive factorisation method and use more sophisticated tools (like QCDf, pQCD, SCET etc.) to compute these decays more accurately \cite{qcd1np}\cite{qcd2np}\cite{qcd3np}\cite{kagan}.
\cng{It has been pointed out in \cite{qcd2np}\cite{kagan}\cite{annihilationdiagram} that a major contributor to transverse amplitudes are the \emph{annihilation diagrams} which can explain the large fraction of transverse amplitudes observed experimentally. }
}

\mycomment{However, since the annihilation diagram contribution to Chromomagnetic operator is a subleading correction to an already subleading term, it is a highly suppressed contribution which can be ignored. Thus, the hierarchy predicted by the $V-A$ structure of current is followed in Chromomagnetic operator.}

\par
\cng{On the other hand, \emph{the contribution from the Chromomagnetic operator is suppressed in transverse penguin amplitudes} (originally pointed out in \cite{kagan}, and verified by pQCD approach in \cite{bvv8}). Therefore, the NP contributions manifesting via Chromomagnetic operator should predominantly contibute to longitudinal polarisation amplitude. This is a key point, that we would use in the subsequent sections for our fit.}
\par
\cng{The total amplitude for SM, left- and right-handed currents can be written as follows, where, as discussed above, we neglect the transverse contributions $\Amp_{\parallel,\perp}^{L,R}$ in NP amplitudes:}
\begin{align}\label{phiphitotalamp}
\begin{aligned}
\Amp^{\rm SM,  Total}_{\phi\phi} &= \Amp_{0,\phi\phi}^{\rm SM} + \Amp_{\parallel,\phi\phi}^{\rm SM} + \Amp_{\perp,\phi\phi}^{\rm SM} \\
&= -\frac{G_F}{\sqrt{2}} V_{tb}^* V_{ts}  (\xi^{\rm SM}_{0}\mx_0^{\rm SM} + \xi^{\rm SM}_{\parallel}\mx_\parallel^{\rm SM} + \xi^{\rm SM}_{\perp}\mx_\perp^{\rm SM} ) \\
\Amp^{\rm L,  Total}_{\phi\phi} &= \Amp_{0,\phi\phi}^{\rm L} \\
&= -\frac{G_F}{\sqrt{2}} V_{tb}^* V_{ts} (\xi^{\rm L}_{0}\mx_0^{\rm NP} ) \\
\Amp^{\rm R,  Total}_{\phi\phi} &= \Amp_{0,\phi\phi}^{\rm R} \\
&= -\frac{G_F}{\sqrt{2}} V_{tb}^* V_{ts} (-\xi^{\rm R}_{0}\mx_0^{\rm NP} ), 
\end{aligned}
\end{align}
$\mx^{\rm SM}$ and $\mx^{\rm NP}$ contains the contribution from the matrix elements for SM and NP case, respectively. The $\xi^p_k$ ($k=\{0,\parallel,\perp\}$ and $p \in \{\rm{SM},\rm{L},\rm{R}\}$) are combinations of Wilson coefficients, and contain the weak phases. \cng{The actual form of $\xi$ and $\mx$ depend upon the model chosen to compute the matrix elements, but it is not important for our purposes. The important thing to notice is the \emph{sign change in the longitudinal component of right-handed amplitude}. This sign change occurs due to the sign change in the axial part of the current; we have verified this for longitudinal amplitude by both naive factorisation and pQCD approach.}

\section{\cng{New phase scheme from Chromomagnetic operator}}\label{sec:newphasescheme}
\cng{
Following Eq.~\eqref{phiphitotalamp}, the total longitudinal transversity amplitude in the presence of NP manifested via Chromomagnetic operator is now given by
\begin{align}\label{transamptotal}
\begin{aligned}
\Amp^{\rm Total}_{0,\phi\phi} &= -\frac{G_F}{\sqrt{2}} V_{tb}^* V_{ts}  (\xi^{\rm SM}_{0}\mx_0^{\rm SM} + \xi^{\rm L}_{0}\mx_0^{\rm NP} - \xi^{\rm R}_{0}\mx_0^{\rm NP} ) \\
&= \Amp_{0,\phi\phi}^{\rm SM} (1 + r^{\rm L} e^{i(\omega_{\rm L} + \sigma)} - r^{\rm R} e^{i(\omega_{\rm R} + \sigma)} ),
\end{aligned}
\end{align}
}
where we parametrise the NP contribution as follows:
\begin{align}\label{npparam}
\frac{\xi^{\rm{L}}_{0} \mx^{\rm NP}_0}{\xi^{\rm{SM}}_{0} \mx^{\rm SM}_0} =  r^{\rm{L}} e^{i (\omega_{\rm{L}}+\sigma)} \quad\quad\quad \frac{\xi^{\rm{R}}_{0} \mx^{\rm NP}_0}{\xi^{\rm{SM}}_{0} \mx^{\rm SM}_0} =  r^{\rm{R}} e^{i (\omega_{\rm{R}}+\sigma)}.
\end{align}
$\omega_{\rm L,R}$ are the weak/CP-odd phases and $\sigma$ is a strong/CP-even phase.
Recalling the definition of interference phase from Eq.~\eqref{intphase} and putting Eq.~\eqref{transamptotal} in it, we can write 
\begin{align}\label{rationalisedphiphi}
 \frac{q}{p} \frac{\bar{\Amp}^{\rm Total}_{0,\phi\phi}}{\Amp^{\rm Total}_{0,\phi\phi}}  = \lambda_0 e^{-i(\theta_0 - \theta_0^c)} = \frac{1 + 2\cos\sigma(r^{\rm L} e^{-i\omega_{\rm L}} + r^{\rm R} e^{-i\omega_{\rm R}} )}{1+2 r^{\rm L} \cos(\omega_{\rm L} + \sigma) - 2r^{\rm R}\cos(\omega_{\rm R} + \sigma)}.
\end{align}

Therefore, only $\lambda_0$ and $\theta_0-\theta_0^c$ would get contributions from NP, while other transversities CP-violating parameters would assume their SM values, i.e $\theta_\parallel = \theta_\parallel^c  = \theta_\perp = \theta_\perp^c =0$ and $\lambda_\parallel = \lambda_\perp = 1$.

As we can see, the five theoretical parameters ($r^{\rm L,R}, \omega_{\rm L,R}$ and $\sigma$) cannot be determined, as we do not have sufficient observables (only $\lambda_0$ and $\theta_0-\theta_0^c$). Nevertheless, an observation of non-zero value of $\lambda_0-1$ and/or $\theta_0 - \theta_0^c$ would clearly indicate the presence of NP. 
\par
Let us now compare the phase scheme that LHCb used in their fit to ours.  Before comparing with our parametrisation, we note that the interference phase in LHCb is defined as
\begin{align}\label{intphaselhcb}
 \frac{q}{p} \frac{\bar{\Amp}^{\rm Total}_{k,\phi\phi}}{\Amp^{\rm Total}_{k,\phi\phi}} = \eta_k\lambda_k e^{-i\phi^{\rm LHCb}_k},
\end{align}
where $k$ is the transversity and $\eta_k$ is the CP parity of the transversity state.  Comparing Eq.~\eqref{intphaselhcb} with Eq.~\eqref{intphase}, we find $\phi^{\rm LHCb}_k \equiv \theta_k - \theta^c_k$. 
LHCb uses the following two different fit configurations:
\begin{itemize}
\item LHCb {\it helicity-dependent} (HD) scheme: \\
$\phi^{\rm LHCb}_{0} = 0$, $\lambda_k = 1$ $\forall k$ ($\phi^{\rm LHCb}_{\perp}$ and $\phi^{\rm LHCb}_{\parallel}$ are the {CP-violating fit} parameters). 
\item LHCb {\it helicity-independent} (HI) scheme: \\
 $\phi=\phi_{k}^{\rm LHCb}$ $\forall k$, $\lambda = \lambda_k$ $\forall k$ ($\phi$ and $\lambda$ are the  {CP-violating fit} parameters). 
\end{itemize}
The new fit configuration we are proposing is
\begin{itemize}
\item  NP manifested via Chromomagnetic operator: \\
$\phi_\perp^{\rm LHCb}$=$\phi_\parallel^{\rm LHCb}$=0 or equivalently $\theta_\parallel = \theta_\parallel^c  = \theta_\perp = \theta_\perp^c =0$, $\lambda_\perp$=$\lambda_\parallel$=1
\\ ($\phi_0^{\rm LHCb}$ and $\lambda_0$ are the  {CP-violating fit} parameters).  
\end{itemize}
The LHCb fit configuration does not match to ours, and a new fit of LHCb data with this new scheme based on our model would be very interesting. \cng{We emphasise that neither of the two LHCb schemes above fit $\phi_0^{\rm LHCb}$ and $\lambda_0$ simultaneously; therefore, our phase scheme is a new avenue to search for NP manifesting itself via Chromomagnetic operator. }

\section{Sensitivity study with the new fit configuration}\label{sec:sensitivitystudy}
In this section, we illustrate an analysis with the new phase scheme we proposed in Section~\ref{sec:newphasescheme}. To start, let us list all the possible fit parameters before considering any model assumptions: $(|A_{0,\perp,\parallel}|^2, \delta_{0,\perp,\parallel}, \theta_{0,\perp,\parallel}, \theta^c_{0,\perp,\parallel}$,$\lambda_{0,\perp,\parallel}$). 
First, using the relation $|A_{0}|^2+|A_{\perp}|^2+|A_{\parallel}|^2=1$, we remove one of the amplitudes, e.g. $|A_{\parallel}|^2$. 
Next, we notice that in Table~\ref{timedeptablestrongCP}, the phases always appear as combinations of $\theta_k-\theta_k^c$ and $\psi_i-\psi_j$ where  $\psi_k\equiv \theta_k+\delta_k$. {For example, the combination $\theta_k^c+\delta_k$  can be rewritten as $\psi_k-(\theta_k-\theta_k^c)$.} Now, let us use {the results of our model} introduced in Section~\ref{sec:npmodel}: Chromomagnetic operator contributes predominantly to longitudinal polarisation, giving us $\theta_\parallel = \theta_\parallel^c  = \theta_\perp = \theta_\perp^c =0$. Then, the arguments of trigonometric functions in Table~\ref{timedeptablestrongCP} can be expressed by the three parameters, ($\theta_0-\theta_0^c, \delta_\parallel-\delta_\perp, \delta_\parallel-\delta_0-\theta_0$). As explained in Section~\ref{cpvnpparam}, the first parameter {is the phase in the interference of decay with and without mixing, which is a CP-violating quantity}, while the last two contain strong phases. Thus, only the first one can be used for a null test.  Finally, our model also imposes $\lambda_{\perp, \parallel}=1$. As a result, we are left with 6 parameters to fit 
\[(\lambda_0, \theta_0-\theta_0^c, \delta_\parallel-\delta_\perp, \delta_\parallel-\delta_0-\theta_0, |A_0|^2, |A_{\perp}|^2)\]
Only the first two can be used for a null test: $\lambda_0\neq1$ and/or $\theta_0-\theta_0^c\neq 0 $ is a clear signal of new physics. \cng{We emphasise that some simplifying assumptions are always going to be required to reduce the number of parameters to achieve a convergent fit, even when more data is collected, since the structure of equations is such that there are more parameters than the number of equations to be solved. }

To illustrate the fit, we first construct two {\it pseudo datasets} by using the LHCb best-fit values, denoted as Data HI and Data HD for the LHCb helicity-independent and helicity-dependent fit, respectively. 
The details of the statistical procedure applied in this study are given in Appendix~\ref{montecarlo}. 
\par
Our fit results are shown in Table~\ref{dcp}, and the correlation matrices are given in Appendix~\ref{corrmat}. 
We note that the results using Data HI and Data HD agree relatively well. 
The obtained uncertainty of $\sigma(\lambda_0)=6-7$\% and $\sigma(\theta_0-\theta_0^c)=5-6$\% with the currently available LHCb statistics (5 $\rm fb^{-1}$)  may be used as an indication for future studies. 

\begin{table}[H]
\begin{center}
\begin{tabular}{|c|c|c|c|c|}
\hline
& \multicolumn{2}{c|}{Data HD} & \multicolumn{2}{c|}{Data HI} \\
\hline
Fit Parameter & Central Value & $\sigma$ & Central Value & $\sigma$ \\
\hline
$\lambda_0$ & 0.978 & 0.058 & 0.984 & 0.070 \\
\hline
$|A_0|^2$ & 0.386 & 0.025 & 0.385 & 0.032 \\
\hline
$|A_{\perp}|^2$ & 0.287 & 0.018 & 0.288 & 0.036 \\
\hline
$\theta_0 - \theta^c_0$ & -0.002 & 0.055 & 0.066 & 0.053 \\
\hline
$\delta_{\parallel} - \delta_{\perp}$ & -0.259 & 0.054 & -0.261 & 0.056 \\
\hline
$\delta_{\parallel} -\delta_{0} - \theta_0$ & 2.560 & 0.071 & 2.589 & 0.079 \\
\hline
\end{tabular}
\caption{Fit results based on our model assumptions, i.e. longitudinal component dominance for NP contributions coming from Chromomagnetic operator ($\theta^c_{\parallel} = \theta_{\parallel} = \theta^c_{\perp} = \theta_{\perp} = 0$ and $\lambda _{\parallel} = \lambda_{\perp} = 1$).}
\label{dcp}
\end{center}
\end{table}

\section{Left or right: \boldmath $B^0_d \rightarrow \phi K_S$ decay}
A decay very similar to $B_s^0 \rightarrow \phi \phi$ decay is the $B_d^0 \rightarrow \phi K_s$ decay, since at the quark level, both contain a $\bar{b} \rightarrow \bar{s} s \bar{s}$ decay. We thus expect the weak interaction to be the same in both the decays, while strong interaction may differ. {In this section, we investigate how the experimental results of $B_d^0 \rightarrow \phi K_s$ complements the $B_s^0 \rightarrow \phi \phi$ results, within the left- and right-handed Chromomagnetic operator model}.  
\par
We start with the $B_d^0 \rightarrow \phi K_s$ decay. The phase in the interference of decay with and without mixing in SM is $2\phi_1$, where $\phi_1$ is the unitary triangle angle. However, NP contributions may deviate its value from $\phi_1$, and what we measure experimentally should then be called $2\phi_1^{\rm eff}$. Using Eq.~\eqref{amptotalphiks}, we can thus write 
\begin{align}
 \frac{q}{p} \frac{\bar{\Amp}^{\rm Total}_{\phi K_s}}{\Amp^{\rm Total}_{\phi K_s}}  = -\lambda_{\phi K_s}e^{-2i\phi_1^{\rm eff}} = -e^{-2i\phi_1} \left( \frac{1 +  \hat{r}^{\rm{L}} e^{i (-\hat{\omega}_{\rm{L}} + \hat{\sigma} )} + \hat{r}^{\rm{R}} e^{i (-\hat{\omega}_{\rm{R}} + \hat{\sigma})}}{1 +  \hat{r}^{\rm{L}} e^{i (\hat{\omega}_{\rm{L}} + \hat{\sigma} )} + \hat{r}^{\rm{R}} e^{i (\hat{\omega}_{\rm{R}} + \hat{\sigma})}} \right),
\end{align}  
{where the negative sign is present as $\phi K_s$ is a CP-odd state}. Rearranging and rationalising the RHS, we get
\begin{align}\label{rationalisedphiks}
\lambda_{\phi K_s}e^{-2i(\phi_1^{\rm eff}-\phi_1)} = \frac{1 + 2\cos\hat{\sigma}(\hat{r}^{\rm L} e^{-i\hat{\omega}_{\rm L}} + \hat{r}^{\rm R} e^{-i\hat{\omega}_{\rm R}} )   }{1 + 2\hat{r}^{\rm L} \cos(\hat{\omega}_{\rm L} + \hat{\sigma}) + 2\hat{r}^{\rm R} \cos(\hat{\omega}_{\rm R} + \hat{\sigma}) }
\end{align} 
We can now compare Eq.~\eqref{rationalisedphiphi} and Eq.~\eqref{rationalisedphiks}. As mentioned before, we assume the weak interaction contribution from NP to be the same for both the decays, thus making $\omega_{\rm L,R} = \hat{\omega}_{\rm L,R}$. In addition, we assume that $r^{\rm L,R}$ and $\hat{r}^{\rm L,R}$ are small and positive. This implies that the sign of the strong interaction (coming from the ratio of matrix elements $\frac{\mx^{\rm NP}_{0}}{\mx^{\rm SM}_{0}}$ and $\frac{\mx^{\rm NP}_{\phi Ks}}{\mx^{\rm SM}_{\phi Ks}}$) is contained in the terms $\cos\sigma$ and $\cos\hat{\sigma}$, respectively. In addition, we see in Eq.~\eqref{rationalisedphiphi} and Eq.~\eqref{rationalisedphiks} that the right-handed contribution from NP has opposite signs for the two cases in the denominator. Therefore, if we can theoretically predict the sign of $\cos\sigma$ and $\cos\hat{\sigma}$ \cng{(which could be done, for example, by pQCD approach \cite{strongphase}),} we can tell the chirality of NP in the following two cases:
\\
\underline{Case 1: Only left-handed NP is present ($r^{\rm R} = \hat{r}^{\rm R} = 0$)}
\\
Taking the ratio of real and imaginary parts of Eq.~\eqref{rationalisedphiphi} and Eq.~\eqref{rationalisedphiks}, and expanding in $r^{\rm L}$ and $\hat{r}^{\rm L}$, we get
\begin{align}
\begin{aligned}
\tan(\theta_0 - \theta_0^c) &\approx 2 r^{\rm L} \sin\omega_{\rm L}\cos\sigma + {\rm O} ((r^{\rm L })^2) \\
\tan(2\phi^{\rm eff}_1 - 2\phi_1) &\approx 2 \hat{r}^{\rm L} \sin\hat{\omega}_{\rm L}\cos\hat{\sigma} + {\rm O} ((\hat{r}^{\rm L })^2)
\end{aligned}
\end{align}
At this point we can define a quantity $\Sigma \equiv [\tan(\theta_0 - \theta_0^c)\tan(2\phi^{\rm eff}_1 - 2\phi_1)]$. As we defined  $r^L$ and $\hat{r}^L$ to be positive and the Chromomagnetic operator leads to $\omega_L=\hat{\omega}_L$, we obtain the relation
\begin{align}
sign(\Sigma)=sign(\cos\sigma\cos\hat{\sigma})
\end{align}
\\
\underline{Case 2: Only right-handed NP is present ($r^{\rm L} = \hat{r}^{\rm L} = 0$)}
\\
Taking the ratio of real and imaginary parts of Eq.~\eqref{rationalisedphiphi} and Eq.~\eqref{rationalisedphiks}, and expanding in $r^{\rm R}$ and $\hat{r}^{\rm R}$, we get
\begin{align}
\begin{aligned}
\tan(\theta_0 - \theta_0^c) &\approx -2 r^{\rm R} \sin\omega_{\rm R}\cos\sigma + {\rm O} ((r^{\rm R })^2) \\
\tan(2\phi^{\rm eff}_1 - 2\phi_1) &\approx 2 \hat{r}^{\rm R} \sin\hat{\omega}_{\rm R}\cos\hat{\sigma} + {\rm O} ((\hat{r}^{\rm R })^2)
\end{aligned}
\end{align}
Thus, in this case, we find an opposite relative sign with respect to the left-handed model:
\begin{align}
sign(\Sigma)=-sign(\cos\sigma\cos\hat{\sigma})
\end{align}
Hence, if the experiments show non-zero CP violating phase results, one can test the chirality of the NP contribution by combining the 
$B_s^0 \rightarrow \phi \phi$ and $B_d^0 \rightarrow \phi K_s$ decay {measurements}, along with the relative sign of $\cos\delta$  and $\cos\hat{\delta}$, which might be obtained theoretically. This conclusion is summarised in Table~\ref{relativesigntable}.
\mycomment{
\par
\cng{Let us look more closely at the viability of sign prediction calculation of the cosine of strong phase. Methods like pQCD or QCDf can predict strong phases, albeit with uncertainties (some constraints can be put from experimental predictions of direct CP asymmetries or direct measurements of strong phases by a fit). Let us make a conservative approximation (or maybe a very ambitious statement in some contexts) and say that the error on the strong phase is around $50\%$. We can sketch this on graph in Figure~\ref{fig:strong_phase}. The shaded area denotes the range of angles where, if the central value of strong phase $\delta$ lies, the sign of $cos\delta$ would not change, given the error is around $50\%$ The shaded area corresponds to about $70\%$ of the total area. If we assume that all the values of the strong phases are equally probable, it would mean that there's a $70\%$ chance that we would be able to determine the sign of strong phases correctly. However, we know that there's a strong tendency of strong phases to lie near 0 or $\pi$ \cite{strongphase0orpi}, which would enhance even more the probability of predicting the sign. However, since this is beyond the scope of our article, we do not go into the calculations of this sign, and entrust this responsibility to the experts of QCD computation.  }

\begin{figure}[t]
\setlength{\unitlength}{1mm}
  \centering
  \begin{picture}(140,60)
    \put(0,-1){
      \includegraphics*[width=140mm]{Fig1}
    }
  \end{picture}
  \caption{\small Range of angles where, if the central value of strong phase $\delta$ lies, the sign of $\cos\delta$ would not change, given the error is about $50\%$.}
\label{fig:strong_phase}
\end{figure}
}
\begin{table}
\begin{center}
\begin{tabular}{|c|c|c|c|}
\hline
 $\cos\sigma$ & $\cos\hat{\sigma}$ & $\Sigma$ & NP chirality  \\ \hline
+ &+ & + & LH \\
+ &-- & + &RH \\
+ & + & -- & RH \\
+ & -- & -- &LH \\
-- & + & -- & LH \\
-- & -- & -- & RH \\
-- & + & + & RH \\
-- & -- & + & LH \\ \hline
\end{tabular}
\caption{Table demonstrating the chirality of NP arising from different combinations of signs of $\cos\sigma$, $\cos\hat{\sigma}$ and $\Sigma \equiv [\tan(\theta_0 - \theta_0^c)\tan(2\phi^{\rm eff}_1 - 2\phi_1)]$, under the assumption that only left-handed or right-handed NP is present. $\sigma$ and $
\hat{\sigma}$ denote the strong phase difference between NP and SM in $B_s^0 \rightarrow \phi \phi$ and $B_d^0 \rightarrow \phi K_s$ decays, respectively (see Eq.~\eqref{npparam}). $\theta_0 - \theta_0^c$ and $2\phi^{\rm eff}_1$ are the phase in the interference of decays with and without mixing in $B_s^0 \rightarrow \phi \phi$ and $B_d^0 \rightarrow \phi K_s$ decays, respectively. $\phi_1$ is the unitary triangle angle.}
\label{relativesigntable}
\end{center}
\end{table}

\section{Conclusions}
In this article, we investigate a new physics search with the CP violation measurements of the $B_s \rightarrow \phi \phi$ decay. The large statistics of the LHCb experiment allows one to perform the time-dependent angular analysis of this decay channel. Such an analysis gives  access to the information of the helicity amplitudes, which are  sensitive to different types of NP effects. In the LHCb analysis, two types of NP scenarios have been investigated, called helicity-dependent and -independent assumptions. 
In this work, we propose a new search scenario based on the NP model induced by the left- and right-handed Chromomagnetic operators, producing a new quark level $b\to s\bar{s}s$ diagram with an extra source of CP violation. Using the fact that the NP coming from this type of operator is dominated by the longitudinal amplitude, we derive a new scheme of phase assumptions which can be tested by the LHCb experiment. 
The same NP effects can manifest itself in the time-dependent CP asymmetry measurement of $B^0_d \rightarrow \phi K_s$ decay. We found that  Belle(II)'s $B^0_d \rightarrow \phi K_s$ decay measurements could complement LHCb's $B^0_s \rightarrow \phi\phi$ measurement to obtain the chirality of NP operator, under the condition that the signs of the strong phases of these decays can be predicted by the theory. 
Finally, we present a sensitivity study of the CP-violating parameters of our proposed model in order to illustrate how the fit can actually be performed. We show that on top of the two CP-violating parameters, there are four extra parameters to be fitted simultaneously: two amplitudes and two  phases. The theoretical predictions for these extra parameters depend heavily on the models describing the strong interaction. On the other hand, a non-zero measurement of the former two CP violating parameters can be interpreted immediately as a signal of NP. Our sensitivity study shows that LHCb with current statistics can determine these two parameters at $5-7\%$  precision. 
 These numbers are obtained using two pseudo datasets and they might not reflect the reality, though, the sensitivities obtained could be used as an indication for future studies. Even though the current measurements do not show a clear signal of NP, further theoretical and experimental efforts would shed more light on these results, and would pave the way for future studies.

\section*{Acknowledgments}
We would like to express our gratitude to Fran\c cois Le Diberder for his careful reading of the manuscript and help in doing the statistical analysis. We would also like to thank Franz Muheim for the helpful correspondence.

\appendix
\section{Angular Conventions}\label{angularconventions}
\cng{In $B^0_s \rightarrow \phi \phi$ decay, since the two $\phi's$ are indistinguishable, we can randomly assign them (and their decay products) the subscripts 1 and 2. $\theta_{1(2)}$ is the angle between the $K^+_{1(2)}$ meson momentum in the $\phi_{1(2)}$ meson rest frame and $\phi_{1(2)}$ meson momentum in $B^0_s$ meson rest frame. Mathematically, we can write it as
\begin{align}
\cos\theta_{1(2)} = \hat{p}^{(\phi_{1(2)})}_{K^+_{1(2)}}.\hat{p}^{(B^0_s)}_{\phi_{1(2)}}
\end{align}
where the notation $\hat{p}^{(x)}_y$ means momentum of particle $y$ in the frame of particle $x$.
The angle $\Phi$, which is the angle between the two decay planes (or between the perpendiculars of the planes), can be defined as follows:
\begin{align}
\begin{aligned}
    \cos\Phi =(\hat{p}_{K^+_1} \times \hat{p}_{K^+_1}).(\hat{p}_{K^+_2} \times \hat{p}_{K^+_2}) \\
    \sin\Phi \hat{z} = [(\hat{p}_{K^+_1} \times \hat{p}_{K^-_1})\times (\hat{p}_{K^+_2} \times \hat{p}_{K^-_2})]
\end{aligned}
\end{align}
where we choose to define the z-direction by the direction of $\phi_1$ momentum \cite{rosner}}

\mycomment{
\section{The Chromomagnetic Dipole operator} \label{cg}
The Chromomagnetic operator for $\bar{b} \rightarrow \bar{s} g$ is given by
\begin{align}
O_{8g} = \frac{g_s}{8\pi^2} m_b \bar{b}_{\alpha} \sigma^{\mu\nu} (1 + \gamma^5) \frac{\lambda^a_{\alpha\beta}}{2}s_{\beta}G^a_{\mu\nu}.
\end{align}
\cng{To obtain a 4-quark operator from this, we attach a quark current through a virtual gluon . Doing some simplifications by using current conservation, we obtain $O_{8g}^4$}
\mycomment{This operator is written only for on-shell gluons. But while dealing with amplitudes, we need matrix elements, and the corresponding one is given by}($X=\phi,K_s$):
\begin{align}
O_{8g}^4 = \frac{\alpha_s}{\pi} \frac{m_b}{q^2} \bar{b}_{\alpha} \gamma^{\mu} \slashed{q} (1+\gamma^5) \frac{ \lambda^a_{\alpha \beta}}{2} s_{\beta} \bar{q}'_{\rho} \gamma_{\mu} \frac{\lambda^a_{\rho \sigma}} {2} q'_{\sigma}.
\end{align}
with $q$ being the gluon momentum. Approximating $q^{\mu} = \sqrt{\langle q^2 \rangle} \frac{p_b^{\mu}}{m_b}$ \footnote{\cng{Note that even without using this approximation, we can write the matrix elements of $O_{8g}^4$ operator in terms of $O_{3-6}$. This approximation just simplifies the calculation, as after doing this calculation without this approximation, we see that the answer has the same structure as given here.} }, and using the Fierz identity $\frac{ \lambda^a_{\alpha \beta}}{2} \frac{\lambda^a_{\rho \sigma}} {2} = \frac{1}{2} ( \delta_{\alpha \sigma} \delta_{\beta\rho} - \frac{1}{3} \delta_{\alpha\beta} \delta_{\rho\sigma}) $ and Dirac equation, we can write the matrix element corresponding to $O_{8g}^4$ operator in terms of $O_{3-6}$ as
\begin{align}
\langle O_{8g}^4 \rangle = \frac{\alpha_s}{4 \pi} \frac{m_b}{\sqrt{\langle q^2 \rangle}} \left[\langle O_4\rangle + \langle O_6 \rangle -  \frac{1}{3}(\langle O_3 \rangle + \langle O_5 \rangle) \right],
\end{align}
where $\langle q^2 \rangle$ is average value of $q^2$, varying between $\frac{m_b^2}{4}$ and $\frac{m_b^2}{2}$ \cite{qrange1}\cite{qrange2}.
\par
Thus, we can write the effective Wilson coefficients for operators $O_{3-6}$ as
\begin{align}
\begin{aligned}
C_3^{\rm{L(R)}} &=   - \frac{1}{3} \frac{\alpha_s}{4\pi} \frac{m_b}{\sqrt{\langle q^2 \rangle}}\ptwiddle{C}_{8g} \\
C_4^{\rm{L(R)}} &=   \frac{\alpha_s}{4\pi} \frac{m_b}{\sqrt{\langle q^2 \rangle}} \ptwiddle{C}_{8g} \\
C_5^{\rm{L(R)}} &=   - \frac{1}{3} \frac{\alpha_s}{4\pi} \frac{m_b}{\sqrt{\langle q^2 \rangle}} \ptwiddle{C}_{8g} \\
C_6^{\rm{L(R)}} &=   \frac{\alpha_s}{4\pi} \frac{m_b}{\sqrt{\langle q^2 \rangle}} \ptwiddle{C}_{8g}.
\end{aligned}
\end{align}
}
\section{Statistical procedure} \label{montecarlo}
The LHCb experimental observables $(a_i, b_i, c_i$ and $d_i)$ are given in Table~\ref{timedeptablestrongCP}: they are the LHCb observables. The only available information from LHCb is the result of fit of those measurements to the theory parameters ($|A_k|^2, \delta_k, \phi_k$), given in~\cite{lhcb}. Therefore, in our study,  we first construct {\it pseudo dataset}, i.e. the central values and the covariance matrices for the LHCb observables,  from this available information. The covariance matrix is obtained by using 
\begin{align}
V_{ij}^{-1} = N \int\  \left.\left(\frac{\partial \hat{f}(x)_{\vec{v}}}{\partial v_i}  \frac{\partial \hat{f}(x)_{\vec{v}}}{\partial v_j}  \frac{1}{\hat{f}(x)_{\vec{v}}}\right)\right|_{\vec{v}=\vec{v}^*}\   dx 
\end{align}
where 
\begin{itemize}
\item $\hat{f}$ is the normalised probability distribution function, which in our case is the angular decay distribution given by Eq.~\eqref{angdist}. Integration over $x$ represents integration over the complete phase space and time. 
\item $\vec{v}$ is the vector of LHCb observables ($a_i, b_i, c_i $ and $ d_i$) that LHCb measures. 
\item $\vec{v}^{\;*}$ is the values of $\vec{v}$ obtained by using the best fit values of the theoretical parameters obtained by LHCb~\cite{lhcb}. Note that there are two fits performed by LHCb with the so-called helicity-independent and -dependent assumptions, and we use both to construct two pseudo datasets. 
\item $N$ is the number of events.
\end{itemize}

Finally, using this {\it pseudo dataset}, we perform a $\chi^2$ fit using $\vec{v}_i$ with our model assumptions, which we call $\vec{v}_{i}^{\rm \; model}$: 
\begin{align}
\chi^2 = \sum_{i,j} (\vec{v}_i^{\rm{ \; model}} - \vec{v}_{i}^{\;*})V^{-1}_{ij}(\vec{v}_j^{\rm{\; model}} - \vec{v}_{j}^{\;*}).
\end{align}

\section{Correlation matrices} \label{corrmat}

\begin{table}[H]
\begin{center}
\begin{tabular}{c|cccccc}
& $\theta_0-\theta^c_0$ & $\delta_{\parallel} - \delta_{\perp}$ & $\delta_{\parallel}-\delta_0-\theta_0$ & $|A_0|^2$ & $|A_{\perp}|^2$ & $\lambda_0$ \\ \hline
$\theta_0-\theta^c_0$ & 1.00 & 0.01 & -0.33 & 0.00 & 0.00 & -0.03 \\
$\delta_{\parallel} - \delta_{\perp}$ & 0.01 & 1.00 & 0.38 & -0.11 & 0.13 & -0.01 \\
$\delta_{\parallel}-\delta_0-\theta_0$ & -0.33 & 0.38 & 1.00 & -0.24 & 0.23 & -0.03 \\
$|A_0|^2$ & 0.00 & -0.11 & -0.24 & 1.00 & -0.72 & -0.67 \\
$|A_{\perp}|^2$ & 0.00 & 0.13 & 0.23 & -0.72 & 1.00 & 0.49 \\
$\lambda_0$ & -0.03 & -0.01 & -0.03 & -0.67 & 0.49 & 1.00 \\
\end{tabular}
\caption{
Correlation matrix based on our model assumptions, i.e. longitudinal component dominance for NP contributions from Chromomagnetic operator ($\theta^c_{\parallel} = \theta_{\parallel} = \theta^c_{\perp} = \theta_{\perp} = 0$ and $\lambda _{\parallel} = \lambda_{\perp} = 1$). Pseudo dataset used: Data HD.}
\end{center}
\end{table}

\begin{table}[H]
\begin{center}
\begin{tabular}{c|cccccc}
& $\theta_0-\theta^c_0$ & $\delta_{\parallel} - \delta_{\perp}$ & $\delta_{\parallel}-\delta_0-\theta_0$ & $|A_0|^2$ & $|A_{\perp}|^2$ & $\lambda_0$ \\ \hline
$\theta_0-\theta^c_0$ & 1.00 & -0.02 & -0.37 & 0.03 & -0.04 & -0.01 \\
$\delta_{\parallel} - \delta_{\perp}$ & -0.02 & 1.00 & 0.40 & -0.06 & 0.07 & -0.04 \\
$\delta_{\parallel}-\delta_0-\theta_0$ & -0.37 & 0.40 & 1.00 & -0.19 & 0.21 & -0.04 \\
$|A_0|^2$ & 0.03 & -0.06 & -0.19 & 1.00 & -0.85 & -0.76 \\
$|A_{\perp}|^2$ & -0.04 & 0.07 & 0.21 & -0.85 & 1.00 & 0.65 \\
$\lambda_0$ & -0.01 & -0.04 & -0.04 & -0.76 & 0.65 & 1.00 \\
\end{tabular}
\caption{Correlation matrix based on our model assumptions, i.e. longitudinal component dominance for NP contributions from Chromomagnetic operator ($\theta^c_{\parallel} = \theta_{\parallel} = \theta^c_{\perp} = \theta_{\perp} = 0$ and $\lambda _{\parallel} = \lambda_{\perp} = 1$). Pseudo dataset used: Data HI.}
\end{center}
\end{table}

\section{\boldmath $B^0_d \rightarrow \phi K_S$ decay}\label{phiksdecay}
The amplitude for $B^0_d \rightarrow \phi K_S$ for SM, left-handed NP and right-handed NP case respectively, can be written as

\begin{align}\label{phiksamp}
\begin{aligned}
\Amp^{\rm{SM}}_{\phi K_s} &= - \frac{G_F}{\sqrt{2}} V^*_{tb} V_{ts} \hat{\xi}^{\rm{SM}}
\mx^{\rm SM}_{\phi K_s} \\
\Amp^{\rm{L}}_{\phi K_s} &= - \frac{G_F}{\sqrt{2}} V^*_{tb} V_{ts} \hat{\xi}^{\rm{L}}
 \mx^{\rm NP}_{\phi K_s} \\
\Amp^{\rm{R}}_{\phi K_s} &= - \frac{G_F}{\sqrt{2}} V^*_{tb} V_{ts} \hat{\xi}^{\rm{R}}
 \mx^{\rm NP}_{\phi K_s},
\end{aligned}
\end{align}
where $ \hat{\xi}^p  (p \in \{\rm{SM},\rm{L},\rm{R}\})$ are  combination of the Wilson coefficients, which contain weak phases, and their exact form depends upon the model chosen to evaluate the matrix elements.
Like for the case of $B^0_s \rightarrow \phi\phi$ decay, the variables $\mx_{\phi K_s}^{\rm SM}$ and $\mx_{\phi K_s}^{\rm NP}$ contain all the information about the matrix elements. Note that ${K}_0$ is a flavour eigenstate, which, by Kaon oscillation,  oscillates between $K_0$ and $\bar{K}_0$ and we see the mass eigenstate $K_S$ in detectors.
\par
Now the total amplitude, which is the sum of all three amplitudes, can be written as
\begin{align}
\Amp^{\rm{Total}}_{\phi K_s} = \Amp^{\rm{SM}}_{\phi K_s} \left( 1 + \frac{\hat{\xi}^{\rm{L}}\mx^{\rm NP}_{\phi K_s}}{\hat{\xi}^{\rm{SM}}\mx^{\rm SM}_{\phi K_s}} + \frac{\hat{\xi}^{\rm{R}}\mx^{\rm NP}_{\phi K_s}}{\hat{\xi}^{\rm{SM}}\mx^{\rm SM}_{\phi K_s}} \right).
\end{align}
 Using similar parametrisation for NP as in Eq.~\eqref{npparam} (but putting hats to differentiate from $B_s^0 \rightarrow \phi \phi$ case), we get
\begin{align}\label{amptotalphiks}
\Amp^{\rm{Total}}_{\phi K_s} = \Amp^{\rm{SM}}_{\phi K_s} \left( 1 +  \hat{r}^{\rm{L}} e^{i (\hat{\omega}_{\rm{L}} + \hat{\sigma} )} + \hat{r}^{\rm{R}} e^{i (\hat{\omega}_{\rm{R}} + \hat{\sigma})} \right).
\end{align}

\section{Coefficients of time-dependent terms}
\cng{The terms in the Table~\ref{timedeptablestrongCP} are the coefficients of time-dependent terms in Eq.~\eqref{timedep}, which are functions of CP-violating parameters. The various quantities used here are defined as follows ($k = \{ \parallel,\perp,0\}$):
\begin{itemize}
    \item $|A_k|$: magnitude of the complete transversity amplitude (see Eqs.~\eqref{ampNP} and \eqref{ampNPconj})
    \item $\delta_k$: strong phase of SM transversity amplitude
    \item $\theta_k$: a mixture of weak and strong phase, as defined in Eqs.~\eqref{ampNP} and \eqref{ampNPconj}, arising due to presence of NP strong and weak phases
\end{itemize}
}
\begin{sidewaystable}
\caption{Coefficients of the time-dependent terms and angular functions used in Eq.~\eqref{timedep}. Amplitudes are defined at $t=0$.}
\resizebox{\textwidth}{!}{
  \begin{tabular}{|c|c|c|c|c|c|c|}
  \hline
i    & $N_i$                               & $a_i$                & $b_i$                             & $c_i $              & $d_i$                             & $f_i $ \\ \hline \hline
1    &$ |A_0|^2$                       & ${(1+\lambda_0^2)}/{2}$       & $-\lambda_0 \cos(\theta^c_0 - \theta_0)$                           &  ${(1-\lambda_0^2)}/{2}$            & $-\lambda_0 \sin(\theta^c_0 - \theta_0)$                             &$4\cos^2\theta_1\cos^2\theta_2$ \\ \hline
2    & $|A_\parallel |^2$           &${(1+\lambda_{\parallel}^2)}/{2}$          &$-\lambda_{\parallel} \cos(\theta^c_{\parallel} - \theta_{\parallel})$             & ${(1-\lambda_{\parallel}^2)}/{2}$              &$-\lambda_{\parallel} \sin(\theta^c_{\parallel} - \theta_{\parallel})$                &$\sin^2\theta_1\sin^2\theta_2(1{+}\cos2\Phi)$ \\ \hline
3    &$ |A_\perp |^2 $              &${(1+\lambda_{\perp}^2)}/{2}$       & $\lambda_{\perp} \cos(\theta^c_{\perp} - \theta_{\perp})$                    &${(1-\lambda_{\perp}^2)}/{2}$            & $\lambda_{\perp} \sin(\theta^c_{\perp} - \theta_{\perp})$                &$\sin^2\theta_1\sin^2\theta_2(1{-}\cos2\Phi)$ \\ \hline
4    &$ {|A_\parallel||A_\perp |}/{2}$   
& $\begin{array}{c} \sin(\delta_{\perp} - \delta_{\parallel} + \theta_{\perp} -\theta_{\parallel} )  \\ -\lambda_{\perp} \lambda_{\parallel} \sin(\delta_{\perp} - \delta_{\parallel} + \theta^c_{\perp} - \theta^c_{\parallel} )  \end{array}$
& $\begin{array}{c} \lambda_{\perp} \sin(\delta_{\perp} - \delta_{\parallel} + \theta^c_{\perp} - \theta_{\parallel} ) \\  - \lambda_{\parallel} \sin(\delta_{\perp} - \delta_{\parallel} + \theta_{\perp} - \theta^c_{\parallel} ) \end{array}$
& $\begin{array}{c} \sin(\delta_{\perp} - \delta_{\parallel} + \theta_{\perp} - \theta_{\parallel})  \\ +\lambda_{\perp} \lambda_{\parallel} \sin(\delta_{\perp} - \delta_{\parallel} + \theta^c_{\perp} - \theta^c_{\parallel} ) \end{array}$
& $\begin{array}{c} -\lambda_{\perp} \cos(\delta_{\perp} - \delta_{\parallel} + \theta^c_{\perp} - \theta_{\parallel} ) \\ -\lambda_{\parallel} \cos(\delta_{\perp} - \delta_{\parallel} + \theta_{\perp} -\theta^c_{\parallel} ) \end{array}$
& $-2\sin^2\theta_1\sin^2\theta_2\sin 2\Phi$ \\ \hline
5       &$ {|A_\parallel||A_0|}/{2} $ 
& $\begin{array}{c} \cos(\delta_0 - \delta_{\parallel} + \theta_{0} - \theta_{\parallel} )  \\ +\lambda_{0} \lambda_{\parallel} \cos(\delta_0 - \delta_{\parallel} + \theta^{c}_{0} - \theta^c_{\parallel} )  \end{array}$
& $ \begin{array}{c} -\lambda_{0} \cos(\delta_0 - \delta_{\parallel} + \theta^c_0 - \theta_{\parallel} ) \\  -\lambda_{\parallel} \cos(\delta_0 - \delta_{\parallel} + \theta_0 - \theta^c_{\parallel} ) \end{array}$
& $\begin{array}{c} \cos(\delta_0 - \delta_{\parallel} + \theta_0 -\theta_{\parallel} )  \\ -\lambda_{0} \lambda_{\parallel} \cos(\delta_0 - \delta_{\parallel} + \theta^c_0 - \theta^c_{\parallel} ) \end{array}$
& $\begin{array}{c} -\lambda_{0} \sin(\delta_0 - \delta_{\parallel} + \theta^c_0 - \theta_{\parallel} ) \\ + \lambda_{\parallel} \sin(\delta_0 - \delta_{\parallel} + \theta_0 - \theta^c_{\parallel} ) \end{array}$
&$ \sqrt{2}\sin2\theta_1\sin2\theta_2\cos\Phi$ \\ \hline
 6       &$ {|A_0||A_\perp |}/{2} $          
& $\begin{array}{c} \sin(\delta_{\perp} - \delta_{0} + \theta_{\perp} -\theta_{0} )  \\ -\lambda_{\perp} \lambda_{0} \sin(\delta_{\perp} - \delta_{0} + \theta^c_{\perp} - \theta^c_{0} )  \end{array}$
& $\begin{array}{c} \lambda_{\perp} \sin(\delta_{\perp} - \delta_{0} + \theta^c_{\perp} - \theta_{0} ) \\  - \lambda_{0} \sin(\delta_{\perp} - \delta_{0} + \theta_{\perp} - \theta^c_{0} ) \end{array}$
& $\begin{array}{c} \sin(\delta_{\perp} - \delta_{0} + \theta_{\perp} - \theta_{0})  \\ +\lambda_{\perp} \lambda_{0} \sin(\delta_{\perp} - \delta_{0} + \theta^c_{\perp} - \theta^c_{0} ) \end{array}$
& $\begin{array}{c} -\lambda_{\perp} \cos(\delta_{\perp} - \delta_{0} + \theta^c_{\perp} - \theta_{0} ) \\ -\lambda_{0} \cos(\delta_{\perp} - \delta_{0} + \theta_{\perp} -\theta^c_{0} ) \end{array}$
& $-\sqrt{2}\sin2\theta_1\sin2\theta_2\sin\Phi $\\\hline
\end{tabular}
}
\label{timedeptablestrongCP}
\end{sidewaystable}

\clearpage

\bibliographystyle{JHEP}
\bibliography{Bibliography.bib}

\providecommand{\href}[2]{#2}\begingroup\raggedright\begin{thebibliography}{10}

\bibitem{cabibbo}
N.~Cabibbo, \emph{{Unitary Symmetry and Leptonic Decays}},
  \href{https://doi.org/10.1103/PhysRevLett.10.531}{\emph{Phys. Rev. Lett.}
  {\bfseries 10} (1963) 531}.

\bibitem{cpvsm}
M.~Kobayashi and T.~Maskawa, \emph{{CP Violation in the Renormalizable Theory
  of Weak Interaction}}, \href{https://doi.org/10.1143/PTP.49.652}{\emph{Prog.
  Theor. Phys.} {\bfseries 49} (1973) 652}.

\bibitem{asymmetrycond}
A.D.~Sakharov, \emph{{Violation of CP Invariance, C asymmetry, and baryon
  asymmetry of the universe}},
  \href{https://doi.org/10.1070/PU1991v034n05ABEH002497}{\emph{Pisma Zh.Eksp.
  Teor. Fiz.} {\bfseries 5} (1967) 32}.

\bibitem{bvv1}
M.~Beneke, J.~Rohrer and D.~Yang, \emph{{Enhanced electroweak penguin amplitude
  in $B \rightarrow VV$ decays}},
  \href{https://doi.org/10.1103/PhysRevLett.96.141801}{\emph{Phys. Rev. Lett.}
  {\bfseries 96} (2006) 141801}
  [\href{https://arxiv.org/abs/hep-ph/0512258}{{\ttfamily hep-ph/0512258}}].

\bibitem{bvv2}
H.-Y.~Cheng and K.-C.~Yang, \emph{{Charmless $B_(s) \rightarrow VV$ decays in
  QCD factorization: Implications of recent $B \rightarrow \phi K^*$
  measurement}},
  \href{https://doi.org/10.1016/S0370-2693(01)00606-2}{\emph{Phys. Lett. B}
  {\bfseries 511} (2001) 40}
  [\href{https://arxiv.org/abs/hep-ph/0104090}{{\ttfamily hep-ph/0104090}}].

\bibitem{bvv11}
R.~Fleischer and M.~Gronau, \emph{{Studying new physics amplitudes in charmless
  $B_(s)$ decays}},
  \href{https://doi.org/10.1016/j.physletb.2007.12.028}{\emph{Phys. Lett. B}
  {\bfseries 660} (2008) 212}
  [\href{https://arxiv.org/abs/0709.4013}{{\ttfamily 0709.4013}}].

\bibitem{bvv12}
A.~Datta, M.~Duraisamy and D.~London, \emph{{New Physics in $b \rightarrow s$
  Transitions and the $B_{d,s}^0 \rightarrow V_1 V_2$ Angular Analysis}},
  \href{https://doi.org/10.1103/PhysRevD.86.076011}{\emph{Phys. Rev. D}
  {\bfseries 86} (2012) 076011}
  [\href{https://arxiv.org/abs/1207.4495}{{\ttfamily 1207.4495}}].

\bibitem{bvv13}
A.~Datta and D.~London, \emph{{Measuring new physics parameters in B penguin
  decays}}, \href{https://doi.org/10.1016/j.physletb.2004.06.069}{\emph{Phys.
  Lett. B} {\bfseries 595} (2004) 453}
  [\href{https://arxiv.org/abs/hep-ph/0404130}{{\ttfamily hep-ph/0404130}}].

\bibitem{bvv14}
A.~Datta and D.~London, \emph{{Triple-product correlations in $B \rightarrow
  V_1 V_2$ decays and new physics}},
  \href{https://doi.org/10.1142/S0217751X04018300}{\emph{Int. J. Mod. Phys. A}
  {\bfseries 19} (2004) 2505}
  [\href{https://arxiv.org/abs/hep-ph/0303159}{{\ttfamily hep-ph/0303159}}].

\bibitem{bvv15}
R.~Fleischer and I.~Dunietz, \emph{{CP violation and CKM phases from angular
  distributions for $B_s$ decays into admixtures of CP eigenstates}},
  \href{https://doi.org/10.1103/PhysRevD.55.259}{\emph{Phys. Rev. D} {\bfseries
  55} (1997) 259} [\href{https://arxiv.org/abs/hep-ph/9605220}{{\ttfamily
  hep-ph/9605220}}].

\bibitem{bvv16}
A.~Datta, M.~Duraisamy and D.~London, \emph{{Searching for New Physics with
  $B$-Decay Fake Triple Products}},
  \href{https://doi.org/10.1016/j.physletb.2011.06.002}{\emph{Phys. Lett. B}
  {\bfseries 701} (2011) 357}
  [\href{https://arxiv.org/abs/1103.2442}{{\ttfamily 1103.2442}}].

\bibitem{bvv17}
I.~Dunietz, R.~Fleischer and U.~Nierste, \emph{{In pursuit of new physics with
  $B_s$ decays}}, \href{https://doi.org/10.1103/PhysRevD.63.114015}{\emph{Phys.
  Rev. D} {\bfseries 63} (2001) 114015}
  [\href{https://arxiv.org/abs/hep-ph/0012219}{{\ttfamily hep-ph/0012219}}].

\bibitem{bvv18}
P.~Ball and R.~Fleischer, \emph{{An Analysis of $B_s$ decays in the left-right
  symmetric model with spontaneous CP violation}},
  \href{https://doi.org/10.1016/S0370-2693(00)00061-7}{\emph{Phys. Lett. B}
  {\bfseries 475} (2000) 111}
  [\href{https://arxiv.org/abs/hep-ph/9912319}{{\ttfamily hep-ph/9912319}}].

\bibitem{bvv19}
R.~Aleksan and L.~Oliver, \emph{{Remarks on the penguin decay $B_s \rightarrow
  \phi \phi$ with prospects for FCCee}},
  \href{https://arxiv.org/abs/2205.07823}{{\ttfamily 2205.07823}}.

\bibitem{bvv20}
D.~London, N.~Sinha and R.~Sinha, \emph{{Bounds on new physics from $B
  \rightarrow V_{(1)} V_{(2)}$ decays}},
  \href{https://doi.org/10.1103/PhysRevD.69.114013}{\emph{Phys. Rev. D}
  {\bfseries 69} (2004) 114013}
  [\href{https://arxiv.org/abs/hep-ph/0402214}{{\ttfamily hep-ph/0402214}}].

\bibitem{qcd1np}
M.~Bartsch, G.~Buchalla and C.~Kraus, \emph{{$B \rightarrow V_L V_L$ decays at
  next-to-leading order in QCD}},
  \href{https://arxiv.org/abs/0810.0249}{{\ttfamily 0810.0249}}.

\bibitem{qcd2np}
M.~Beneke, J.~Rohrer and D.~Yang, \emph{{Branching fractions, polarisation and
  asymmetries of $B \rightarrow VV$ decays}},
  \href{https://doi.org/10.1016/j.nuclphysb.2007.03.020}{\emph{Nucl. Phys. B}
  {\bfseries 774} (2007) 64}
  [\href{https://arxiv.org/abs/hep-ph/0612290}{{\ttfamily hep-ph/0612290}}].

\bibitem{qcd3np}
H.-Y.~Cheng and C.-K.~Chua, \emph{{{QCD} Factorization for Charmless Hadronic
  $B_s$ Decays Revisited}},
  \href{https://doi.org/10.1103/PhysRevD.80.114026}{\emph{Phys. Rev. D}
  {\bfseries 80} (2009) 114026}
  [\href{https://arxiv.org/abs/0910.5237}{{\ttfamily 0910.5237}}].

\bibitem{lhcb}
{\scshape LHCb} collaboration, \emph{{Measurement of CP violation in the $
  {B}_s^0 \rightarrow \phi \phi$ decay and search for the $ B^0\rightarrow
  \phi\phi $ decay}},
  \href{https://doi.org/10.1007/JHEP12(2019)155}{\emph{JHEP} {\bfseries 12}
  (2019) 155} [\href{https://arxiv.org/abs/1907.10003}{{\ttfamily
  1907.10003}}].

\bibitem{hflav}
{\scshape HFLAV} collaboration, \emph{{Averages of $b$-hadron, $c$-hadron, and
  $\tau$-lepton properties as of 2021}},
  \href{https://arxiv.org/abs/2206.07501}{{\ttfamily 2206.07501}}.

\bibitem{transversity}
I.~Dunietz, H.R.~Quinn, A.~Snyder, W.~Toki and H.J.~Lipkin, \emph{{How to
  extract CP violating asymmetries from angular correlations}},
  \href{https://doi.org/10.1103/PhysRevD.43.2193}{\emph{Phys. Rev. D}
  {\bfseries 43} (1991) 2193}.

\bibitem{oscfreq}
{\scshape LHCb} collaboration, \emph{{Precision measurement of the
  $B^{0}_{s}$-$\bar{B}^{0}_{s}$ oscillation frequency with the decay
  $B^{0}_{s}\rightarrow D^{-}_{s}\pi^{+}$}},
  \href{https://doi.org/10.1088/1367-2630/15/5/053021}{\emph{New J. Phys.}
  {\bfseries 15} (2013) 053021}
  [\href{https://arxiv.org/abs/1304.4741}{{\ttfamily 1304.4741}}].

\bibitem{angdist}
B.~Bhattacharya, A.~Datta, M.~Duraisamy and D.~London, \emph{{Searching for new
  physics with $B_s^0\rightarrow V_1V_2$ penguin decays}},
  \href{https://doi.org/10.1103/PhysRevD.88.016007}{\emph{Phys.Rev.} {\bfseries
  D88} (2013) 016007} [\href{https://arxiv.org/abs/1306.1911}{{\ttfamily
  1306.1911}}].

\bibitem{rescattering}
H.-Y.~Cheng, C.-K.~Chua and A.~Soni, \emph{{Final state interactions in
  hadronic B decays}},
  \href{https://doi.org/10.1103/PhysRevD.71.014030}{\emph{Phys. Rev. D}
  {\bfseries 71} (2005) 014030}
  [\href{https://arxiv.org/abs/hep-ph/0409317}{{\ttfamily hep-ph/0409317}}].

\bibitem{mixingcpasymmetry}
{\scshape LHCb} collaboration, \emph{{Measurement of the $CP$ asymmetry in
  $B_s^0-\bar{B}_s^0$ mixing}},
  \href{https://doi.org/10.1103/PhysRevLett.117.061803}{\emph{Phys. Rev. Lett.}
  {\bfseries 117} (2016) 061803}
  [\href{https://arxiv.org/abs/1605.09768}{{\ttfamily 1605.09768}}].

\bibitem{rosner}
M.~Gronau and J.L.~Rosner, \emph{{Triple product asymmetries in $K$, $D_{(s)}$
  and $B_{(s)}$ decays}},
  \href{https://doi.org/10.1103/PhysRevD.84.096013}{\emph{Phys. Rev. D}
  {\bfseries 84} (2011) 096013}
  [\href{https://arxiv.org/abs/1107.1232}{{\ttfamily 1107.1232}}].

\bibitem{bigisandracpv}
I.~Bigi and A.~Sanda, \emph{CP Violation}, Cambridge Monographs on Particle
  Physics, Nuclear Physics and Cosmology, Cambridge University Press (2000).

\bibitem{pnp1}
Y.Y.~Keum, \emph{{New physics search in B meson decays}},
  \href{https://doi.org/10.1142/9789812791870_0020}{\emph{{3rd International
  Conference on B Physics and CP Violation (BCONF99)}} (1999) 140}
  [\href{https://arxiv.org/abs/hep-ph/0003155}{{\ttfamily hep-ph/0003155}}].

\bibitem{pnp2}
T.~Moroi, \emph{{CP violation in $B_d \rightarrow \phi K_s$ in SUSY GUT with
  right-handed neutrinos}},
  \href{https://doi.org/10.1016/S0370-2693(00)01160-6}{\emph{Phys. Lett. B}
  {\bfseries 493} (2000) 366}
  [\href{https://arxiv.org/abs/hep-ph/0007328}{{\ttfamily hep-ph/0007328}}].

\bibitem{pnp3}
M.-B.~Causse, \emph{{Supersymmetric penguin contributions to the process $B_d
  \rightarrow \phi K_s$ in SUSY GUT with right-handed neutrino}},
  \href{https://arxiv.org/abs/hep-ph/0207070}{{\ttfamily hep-ph/0207070}}.

\bibitem{pnp4}
G.L.~Kane, P.~Ko, H.-B.~Wang, C.~Kolda, J.-H.~Park and L.-T.~Wang, \emph{{$B_d
  \rightarrow \phi K_{s}$ and supersymmetry}},
  \href{https://doi.org/10.1103/PhysRevD.70.035015}{\emph{Phys. Rev. D}
  {\bfseries 70} (2004) 035015}
  [\href{https://arxiv.org/abs/hep-ph/0212092}{{\ttfamily hep-ph/0212092}}].

\bibitem{pnp5}
R.~Harnik, D.T.~Larson, H.~Murayama and A.~Pierce, \emph{{Atmospheric neutrinos
  can make beauty strange}},
  \href{https://doi.org/10.1103/PhysRevD.69.094024}{\emph{Phys. Rev. D}
  {\bfseries 69} (2004) 094024}
  [\href{https://arxiv.org/abs/hep-ph/0212180}{{\ttfamily hep-ph/0212180}}].

\bibitem{pnp6}
S.~Baek, \emph{{CP violation in $B \rightarrow \phi K_s$ decay at large
  $\tan\beta$}}, \href{https://doi.org/10.1103/PhysRevD.67.096004}{\emph{Phys.
  Rev. D} {\bfseries 67} (2003) 096004}
  [\href{https://arxiv.org/abs/hep-ph/0301269}{{\ttfamily hep-ph/0301269}}].

\bibitem{pnp7}
S.~Khalil and E.~Kou, \emph{{A Possible supersymmetric solution to the
  discrepancy between $B \rightarrow \phi K_s$ and $B \rightarrow \eta'
  K_{(s)}$ CP asymmetries}},
  \href{https://doi.org/10.1103/PhysRevLett.91.241602}{\emph{Phys. Rev. Lett.}
  {\bfseries 91} (2003) 241602}
  [\href{https://arxiv.org/abs/hep-ph/0303214}{{\ttfamily hep-ph/0303214}}].

\bibitem{pnp8}
M.~Raidal, \emph{{CP asymmetry in $B \rightarrow \phi K_s$ decays in left-right
  models and its implications on $B_s$ decays}},
  \href{https://doi.org/10.1103/PhysRevLett.89.231803}{\emph{Phys. Rev. Lett.}
  {\bfseries 89} (2002) 231803}
  [\href{https://arxiv.org/abs/hep-ph/0208091}{{\ttfamily hep-ph/0208091}}].

\bibitem{cpvinosc}
{\scshape LHCb} collaboration, \emph{{Implications of LHCb measurements and
  future prospects}},
  \href{https://doi.org/10.1140/epjc/s10052-013-2373-2}{\emph{Eur. Phys. J. C}
  {\bfseries 73} (2013) 2373}
  [\href{https://arxiv.org/abs/1208.3355}{{\ttfamily 1208.3355}}].

\bibitem{lhcbjpsimixing}
{\scshape LHCb} collaboration, \emph{{Updated measurement of time-dependent
  CP-violating observables in $B^{0}_{s}\rightarrow J/\psi K^+ K^-$ decays}},
  \href{https://doi.org/10.1140/epjc/s10052-019-7159-8}{\emph{Eur. Phys. J. C}
  {\bfseries 79} (2019) 706}
  [\href{https://arxiv.org/abs/1906.08356}{{\ttfamily 1906.08356}}].

\bibitem{helampkramer}
G.~Kramer and W.F.~Palmer, \emph{{Branching ratios and CP asymmetries in the
  decay $B \rightarrow V V$}},
  \href{https://doi.org/10.1103/PhysRevD.45.193}{\emph{Phys. Rev. D} {\bfseries
  45} (1992) 193}.

\bibitem{helicityhierarchy1}
J.G.~Korner and G.R.~Goldstein, \emph{{Quark and Particle Helicities in
  Hadronic Charmed Particle Decays}},
  \href{https://doi.org/10.1016/0370-2693(79)90085-6}{\emph{Phys. Lett. B}
  {\bfseries 89} (1979) 105}.

\bibitem{helicityhierarchy2}
A.~Ali, J.G.~Korner, G.~Kramer and J.~Willrodt, \emph{{Nonleptonic Weak Decays
  of Bottom Mesons}}, \href{https://doi.org/10.1007/BF01440227}{\emph{Z. Phys.
  C} {\bfseries 1} (1979) 269}.

\bibitem{kagan}
A.L.~Kagan, \emph{{Polarization in $B \rightarrow VV$ decays}},
  \href{https://doi.org/10.1016/j.physletb.2004.09.030}{\emph{Phys. Lett. B}
  {\bfseries 601} (2004) 151}
  [\href{https://arxiv.org/abs/hep-ph/0405134}{{\ttfamily hep-ph/0405134}}].

\bibitem{phikstar}
{\scshape LHCb} collaboration, \emph{{Measurement of polarization amplitudes
  and CP asymmetries in $B^0 \rightarrow \phi K^*(892)^0$}},
  \href{https://doi.org/10.1007/JHEP05(2014)069}{\emph{JHEP} {\bfseries 05}
  (2014) 069} [\href{https://arxiv.org/abs/1403.2888}{{\ttfamily 1403.2888}}].

\bibitem{annihilationdiagram}
Z.-T.~Zou, A.~Ali, C.-D.~Lu, X.~Liu and Y.~Li, \emph{{Improved Estimates of The
  $B_{(s)}\rightarrow V V$ Decays in Perturbative QCD Approach}},
  \href{https://doi.org/10.1103/PhysRevD.91.054033}{\emph{Phys. Rev. D}
  {\bfseries 91} (2015) 054033}
  [\href{https://arxiv.org/abs/1501.00784}{{\ttfamily 1501.00784}}].

\bibitem{bvv8}
D.-C.~Yan, X.~Liu and Z.-J.~Xiao, \emph{{Anatomy of $B_s \rightarrow VV$ decays
  and effects of next-to-leading order contributions in the perturbative QCD
  factorization approach}},
  \href{https://doi.org/10.1016/j.nuclphysb.2018.08.002}{\emph{Nucl. Phys. B}
  {\bfseries 935} (2018) 17}
  [\href{https://arxiv.org/abs/1807.00606}{{\ttfamily 1807.00606}}].

\bibitem{strongphase}
S.~Mishima and A.I.~Sanda, \emph{{Calculation of magnetic penguin amplitudes in
  $B \rightarrow \phi K$ decays using PQCD approach}},
  \href{https://doi.org/10.1143/PTP.110.549}{\emph{Prog. Theor. Phys.}
  {\bfseries 110} (2003) 549}
  [\href{https://arxiv.org/abs/hep-ph/0305073}{{\ttfamily hep-ph/0305073}}].

\end{thebibliography}\endgroup
\clearpage
 
\end{document}